# Attilio Sacripanti

# *A Biomechanical Reassessment of the Scientific Foundations of Jigorō Kanō's Kōdōkan Jūdō*

extended version with Physical and Mathematical framework

# *A Biomechanical Reassessment of the Scientific Foundations of Jigorō Kanō's Kōdōkan Jūdō*

by *Attilio Sacripanti*\*†‡§\*\*

\*ENEA (National Agency for Environment Technological Innovation and Energy) Robotic Laboratory
†University of Rome II "Tor Vergata", Italy
‡FIJLKAM Italian Judo Wrestling and Karate Federation
§European Judo Union Knowledge Commission Commissioner
\*\*European Judo Union Education Commission Scientific Consultant

## Abstract

In this paper we provide an appraisal of the scientific foundations of the Olympic sport "jūdō" from a Western perspective, *i.e.* a biomechanical reassessment of the basic teaching tools of the foundations of Jigorō Kanō's Kōdōkan jūdō. The core of jūdō functional organization is the triad made up by the "*Kuzushi-Tsukuri-Kake*" movement. *Kuzushi* (unbalance), *tsukuri* (entry and proper fitting of *Tori*'s [the actor] body into the position prior to the throwing phase), and *kake* (execution of the jūdō throwing action) occur as one effective movement without separation. Biomechanical analysis is able to broaden and deepen these steps proposed by Kanō, who for didactic reasons had separated the whole movement into three steps, as was pointed out previously by old Japanese biomechanical studies (1972 -1978). The results of our biomechanical reassessment support:

1. The importance of broadening the unbalance concept to the dynamic situation.
2. Singling out the existence of two classes of paths and trajectories called "Action Invariants". The first, "General Action Invariants", are connected to the whole body motion for shortening the distance in the *kuzushi/tsukuri* phase; the second one, the "Specific Action Invariants", are connected to the superior and inferior kinetic chains motion and right positioning connected to both the *kuzushi* and *tsukuri* phases.
3. The existence of two basic physical principles of throwing in according to a new classification.
4. In a clear way the basic mechanical steps of all throws, and also show hidden connections, similarities and differences among jūdō throwing techniques, making this reassessment particularly useful from a didactic point of view.
5. Demonstrate that this biomechanical approach is also able to understand how new and innovative technical variations are born.

Our reassessment does not reject the educational value of the whole jūdō structure as an educational tool as conceived by Jigorō Kanō-*sensei* now almost 150 years ago, but it helps clarifying his concept and facilitates teachers and coaches in their professional work.

*Attilio Sacripanti*

# *A Biomechanical Reassessment of the Scientific Foundations of Jigorō Kanō's Kōdōkan Jūdō*

# A Biomechanical Reassessment of the Scientific Foundations of Jigorō Kanō's Kōdōkan Jūdō

## 1. The Foundation of Kōdōkan Jūdō

*"Nothing under the sun is greater than education. By educating one person and sending him into the society of his generation, we make a contribution extending a hundred generations to come." Jigorō Kanō statement made at the Kodokan's 50th anniversary in 1934* [1]

***Jigorō Kanō: A daring theory: Using motor development as a basis and tool for education***

Jigorō Kanō was born in 1860 in Mikage in Hyōgo Prefecture and died in 1938. He was the founder of jūdō, but principally he was a great educator and as an educator he always considered jūdō as a training for life, and encouraged his followers to balance both physical and mental aspects of training. In addition to founding *Kōdōkan jūdō*, Kanō also left an impressive legacy as both the Japanese "father of physical education" and the "father of education". At the age of 23, while a teacher at the *Gakushū'in*, he also opened (in addition to the *Kōdōkan dōjō*) the *Kanō Jūku* tutoring school and the *Kōbun Gaku'in* school, for the purpose of providing a well-rounded education in which the physical, mental, and moral aspects are well balanced. Kanō proposed an innovative theory: education based on motor development. Kanō's idea of motor development was very far from simple limbs movement, or from today's wellness concept. His ideas went deeper and were more complex; during his study of old *jūjutsu* techniques he was particularly drawn to the harmony and inner rationality of these techniques, both because of their movements' dynamics and their relationship with the human body. He concluded that this kind of harmony and rationality could be useful and profitable for the education of youngsters.

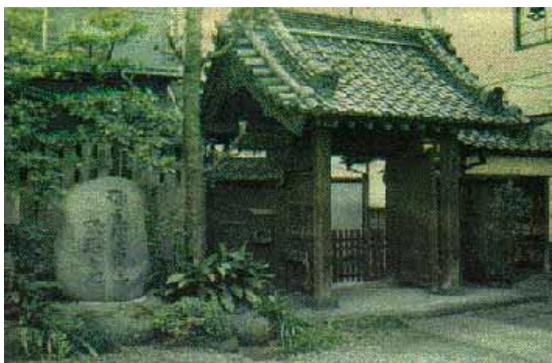 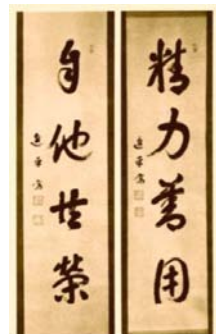

**Fig. 1: *Birthplace of Kōdōkan, Eishō-ji Temple (永昌寺) in Tōkyō right "Sei-ryoku zen'yō" (maximum efficient use of energy), left "Jita kyō-ei" (mutual prosperity for oneself and others) written by Professor Kanō***

Connecting the previous aspects with Confucianist principles, Kanō organized the three aspects of his educational approach (physical, intellectual and moral) into a new educational method: "*Jū-Dō*". Then, starting from *jūjutsu* techniques (utilized for self-protection) he established the jūdō techniques (utilized for self-perfection). This innovative point of view was very important in the history of Japanese education and today is part of local educational theory.

With the help of his first nine disciples Kanō founded his training hall, named "*Kō-dō-kan*", or "the place to teach the way" [3]. Integrating what he considered the positive points of the *Tenjin Shinyō-ryū* and *Kitō-ryū* styles of classical *jūjutsu*, with his own ideas and inspirations, he established a revised body of physical techniques. In addition, he also transformed the traditional *jūjutsu* principle of "defeating strength through flexibility" into a new principle of "maximum efficient use of physical and mental energy." The result was a new theoretical and technical system that Kanō felt, would better match the needs of modern people. The essence of this system he expressed in the axiom "maximum efficient use of energy", a concept he considered both a cornerstone of martial arts and a principle useful in many aspects of life. From the very beginning this obvious pedagogical organization could be recognized in his subdivision of jūdō in three areas: *Rentai-hō*, *Shōbu-hō*, and *Shūshin-hō*. *Rentai-hō* refers to jūdō as a physical exercise, while *Shōbu-ho* is jūdō as a martial art, and *Shushin-hō* is the cultivation of wisdom and virtue as well as the study and application of the principles of jūdō in our daily lives [3].

## 2. The Kernel of Jūdō Teaching

*Kōdōkan jūdō* distinguished itself from the old *jūjutsu* schools, by being organized as a modern teaching structure underpinned by scientific and fundamentals concepts, such as: body position (*shisei*), grips (*kumi-kata*), body's movements (*shintai*). Jūdō's throwing techniques (Nage-waza) were carefully selected from among old jūjutsu techniques or were newly devised (invented by Kanō and/or his students). These throwing techniques were then categorized into five classes (*Go-kyō*) according to the tools that needed to be mobilized to apply the necessary force throw: arm (*Te-waza*), leg (*Ashi-waza*), hips (*Koshi-waza*), or one's own falling body (*Ma & Yoko-sutemi-waza*). Another crucial part was represented by breakfalls (*Ukemi*) meant to how handling one's own body safely and without injury during falls or when being thrown.

The structure of jūdō was based on a progressing scientific understanding. In *Jūdō Kyōhan* (French Edition, 1911)[2], its authors, Yokoyama and Ōshima, speak unambiguously about how jūdō throws work according to the laws of physiology, mechanics, and psychology. In fact, Yokoyama and Ōshima, in their transcription of Professor Kanō's lessons in *Jūdō Kyōhan* write: *"About physical training, Judo is important, because it also gives the technical ability to fight to the developing body, …about fighting techniques, judo is again superior because every part of body works in agreement to the physiological laws, the application of forces agrees with the principles of mechanics, …finally the mind works in agreement with psychology"*. (p.5)

In one chapter about breaking the body's balance Yokoyama and Ōshima write again: *"A lot of erudite mathematical evidence can be introduced to explain this problem … but we prefer to do so using the simple example of a stick motion in space..."* (p.6 ) [2]

These statements support that science, simple proto-biomechanics, underpinned that was practiced and taught by jūdō's founder to his first students. But the inner kernel of teaching and understanding jūdō's throwing techniques relies on a triad as structured by Kanō: ***Kuzushi, Tsukuri, Kake.***

*Kuzushi*, *tsukuri*, and *kake* represent the fundamental building blocks of Nage-waza. Kanō and his assistants relying on their grasp and evolution of their biomechanical analyses *ante litteram*, recognized these three different phases (*kuzushi, tsukuri, kake*) in the dynamics of throwing and in static actions:. The concise *Kōdōkan* dictionary of jūdō defines these terms as follows:

***Kuzushi: Balance-breaking*** → *An action to unbalance your opponent in preparation to throwing him/her.*
***Tsukuri: Positioning set-up*** → *An action to set up a throw after breaking your opponent's balance.*
***Kake: Application, Execution*** → *An action used to execute a technique such as a throw after breaking your opponent's balance (kuzushi) and provoking him into a disadvantageous position (tsukuri)* [3].

A proper and coordinated application of these concepts permits an efficacious way of throwing a rival down on the tatami. There is not a true time wise between each of these phases; they are merely an artificial distinction that serves the pedagogical explication of a throw to facilitate it being taught as a single, fluent and continuous movement. The controversy about any potential priority of either of these phases was settled using electromyographic verification of jūdō competitors' muscular actions. Using this approach some of the older *Kōdōkan* scientific research able to demonstrate that any two phases are both connected [4]. For this reason we can conclude that, for example, the two phases, *kuzushi/tsukuri*, are in fact exactly a single continuous fluent movement starting at the same time without any prioritization of either. However, this subdivision in three phases remains a very useful teaching tool and 'modern' to break down a complex matter. Translated into modern scientific language, this is the application of differential analysis to the entire throwing movement (*i.e.*, to dissect a complex movement into more understandable conceptual sub-steps).

## 2.1. Kuzushi

In the following paragraph we will discuss breaking balance taking into consideration Jigorō Kanō's most important and rudimentary views about this concept and Kazuzō Kudō's expanded approach. We will closely examine these views and the applicable circumstances both from a biomechanical and physical point of view.

Kanō, who was a student of an ancient *jūjutsu* school —*Kitō-ryū,* during training of *randori* (*midare-dori*) with his master Tsunetoshi Iikubo, realized the importance of the concept of *breaking one's opponent's balance* to effect better throwing techniques and a more efficacious execution of these techniques. This is particularly so between two fighters with similar physique and force. In such cases breaking the balance of one's adversary is crucial to obtain results. Old-school Japanese *jūjutsu* consists of using to your advantage the condition of the body which has lost equilibrium. It is called *kuzure-no-jōtai* (state of broken balance). Sometimes the opponent himself loses balance, and at other times you positively destroy the opponent's balance, leading him to a vulnerable posture.

In jūdō, each technique is subdivided and analyzed into *tsukuri* (preparatory action) and *kake*

(attack). Preparatory action is further divided into *aite-no-tsukuri* (preparing of the opponent) and *jibun-no-tsukuri* (preparing oneself). Preparing of the opponent consists of destroying the opponent's balance before performing a technique and putting him in a posture that facilitates application of a technique. At the same instant the one acting must be in a posture and position in which it is easy to apply a technique. This is the "preparing of oneself".

These above comments about the concept of breaking balance are elaborated in the principle of "***Roppō-no-kuzushi***" ("six fundamental directions of unbalance" [2]) which in Kōdōkan jūdō was expanded to **"*Happō-no-kuzushi*"** ("eight fundamental straight directions of unbalance" [5]). The word *kuzushi* (unbalance) was associated by Kanō with the concept of:

*"<u>maximum power in a throwing technique with minimum muscular energy</u>".*

This notion was already extensively developed in the *Kitō-ryū jūjutsu*-school almost as an anticipation of *Kōdōkan jūdō*. One must consider though that given the specific dynamic phases which compose a fight, the problem that surrounding causing an opponent's imbalance evidently is more complex and articulate than the imbalance of a passive, still body.

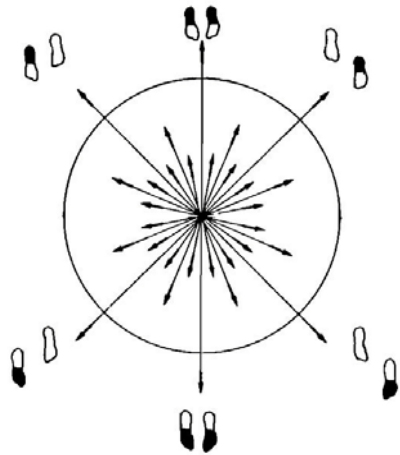

*Fig. 2: Roppō-no-kuzushi*

In fact, the biodynamic solution of such questions needs meticulous study involving: the forces in action, the right direction to unbalance an opponent's body, the proper moving to prepare effective throwing techniques, a way to establish grips, changes in direction and speed of the moving bodies, *etc*.

In a first attempt, Jigorō Kanō experimented with the *go-kyō,* always using an elementary scientific approach. First he would throw static opponents (static phase) in *uchi-komi*, and then he would study the principles of throwing during a dynamic movement while walking in a straight line as we can notice in *nage- no- kata* (dynamic phase) [2]. Kanō used *kuzushi* to synthesize the transfer of action in biomechanical elements, *i.e.* the barycentric projection of the opponent's body outside both the optimal trapezoid surface and the supporting base. The action of *kuzushi* takes advantage of a human body's erect position, "<u>a physic position of unstable equilibrium</u>", by introducing everything that permits throwing that body after an unbalance, namely, through an appropriate barrier (*e.g.*, *tai-otoshi, hiza-guruma*) or through merely causing a hyper-extension of the body in a space (*e.g.*, *uki-otoshi, sumi-otoshi).*
The nicety of jūdō techniques does not lie not in the action of performing techniques, but rather in the skill with which the anticipating work, *i.e.* the preparatory phase, is completed. It was the transparency and the originality of the founder of *Kōdōkan jūdō*'s idea to analyze and clarify how throwing techniques performed in a fraction of a second rely on the practice and study of their preparatory action. In order to "prepare an opponent" for being thrown in the most efficient way, theoretical and practical study about the principles of breaking balance is essential. In addition, to properly master the "preparing of oneself", it is necessary to study the importance of natural posture and the theory and practice of *ma-ai* (proper distance).

The question about the direction of forces and their proper use prompted Kanō to consider the development of dynamic actions with a purely straight and two-dimensional symmetry: *"…if a rival pushes, you must pull in the same direction; if he pulls, you push him in the same direction"* [2]. Kanō and his assistants, in fact, affirmed the principle of maximum efficiency which in biomechanics terminology means "the principle of minimum application of muscular energy". The concept of linearity, a fundamental aspect in the beginning years of jūdō, formed the basis for the system of *Happō-no-kuzushi* or "Eight basic directions of unbalance". All these directions represent a linear and straight symmetry in a biomechanical plane parallel to the *tatami*. In more recent days, one of Kanō's last pupils, Kazuzō Kudō-*sensei*, decided to expand the directional principle of *kuzushi,* to include fourteen fundamental directions of unbalance with identical linear symmetry: *"I would like if you considered my principle about fourteen directions to unbalance the rival, thinking it like a going over 'of the old rudiments': this new principle arises from evolution of technique and from more and more mastery, in a throwing, of many masters."* (pag.36) [6]

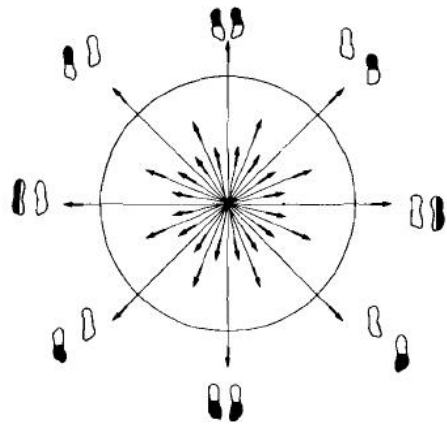

*Fig. 3: Happō-no-kuzushi*

This effort, from a scientific point of view, unfortunately has limited use due to its geometric limitations. Even if we only consider two-dimensional linear symmetry, and because of its complexity exclude specific three-dimensional issues, then the number of unbalance directions is in infinity, more precisely, *"an infinity of the power of continuum"* [8, 9]. The **Happō-no-kuzushi** directional principle is therefore only satisfactory on the condition that each of the eight fundamental and straight directions proposed by Kanō is considered as "*representative vector*" of a group or more exactly of a *class of directions.* We must therefore approach *Happō-no-kuzushi* as a didactic example depicting what is essentially an 'innumerable' number of horizontal straight directions of unbalance now divided into eight classes. In this way, some order is created which allows us to better understand the infinite horizontal directions which an unbalanced erect human body placed on a horizontal plane is subjected to.

## 2.1.1 Evolution in Effectiveness: Rotational Approach

Jigoro Kano wasn't able to develop his research in a rotational field, but he paved the way to permit a natural Judo evolution with the *Itsusu No Kata*.

Kano's scientific method was influenced by Ueshiba's Aikido, in fact Judo acquired, developed and adapted the rotational unbalance concept.

*Kusushi Tsukuri* phase became mainly rotary (but not officially into the books), taking a correct practical analysis of the throwing movement during competition into consideration.

In fact, Kyuzo Mifune (10°dan) used to assert : *"if the rival push you needs to rotate your body; if he pulls, you needs to shift against him in diagonal direction"* [34]

The rotational unbalance is very important to single out the importance of **Tai Sabaki** (体捌き Litt:, body shifting; body control) [3] which must be considered, in open mind, in a most general view. In fact, the rotation, either in attack or in defense, is the base of an effective advanced Judo. **Tai Sabaki**,(体捌き) includes the whole Tori body's movements which will produce a rotational *Kuzushi-Tsukuri* phase.

Mister Koizumi (8°dan), in fact, professed: *"the action to throw should be a continuous curved line..."*[35].

From didactics point of view it is possible to define an attack **Tai Sabaki** and a defense **Tai Sabaki.**.

The unbalance directions, which are tangents to the circle developed, are infinite, like the previous rectilinear *Kuzushi,* too.  Fig.4

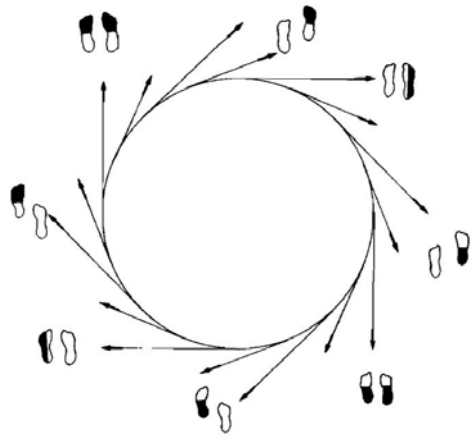

*Fig.4  Rotational Unbalances*

A very little rotational moving, just a few degrees, permits to avoid often, with its deviation, the rival attack force, preventing each other throwing action (defensive **Tai Sabaki**); it's possible, in a more dynamic phase, to benefit by the insufficient rival *kuzushi- tsukuri* action and then utilize one's own unbalance to realize a combined effect between one's own kinetic energy and rival rotational deviation, in such way you can apply a throws (attacking **Tai Sabaki**).

From the biomechanical point of view these subdivisions are useless cause they are too much metaphysical; the essence of the movement is, above all, the body's à rotation, using the fore-foot as pivot.

The introduction and execution of rotational **Kuzushi-Tsukuri** is the practical evolution and presupposes a condition of highly dynamic competition. The agonistic Judo recently uses these rotational concepts as the most natural and appropriate to minimize the efforts and the energy, during the higher dynamic phases, in a modern competition.

### *2.2. Tsukuri*

The word *tsukuri* (positioning set-up) refers to the whole of the unlimited movements which the *jūdōka* can evoke to prepare the realization of a throwing technique. Oftentimes in jūdō books this step is mistranslated as *fitting in* movements, or it is described relative and to each specific throw in the *go-kyō*. In such a way, it is very difficult to find a general description of the *tsukuri* phase as part of the essence of jūdō teaching. One can find one of the best possible definitions of *tsukuri* in the 40-year old book *Dynamic Judo* by Kazuzō Kudō:

*"As we have explained several times, to apply a technique to your opponent you must move together with him and push and pull him in such a way that you force him into a position in which your attack is easy to make and in which he is easily to throw. This is what we call the preparatory moves or in Japanese, the Tsukuri."* (pag 36) [6]

But there is also a very interesting addendum to this otherwise clear explanation, namely, a reference to what essentially represents the inner concept of a "couple system": *"In order to apply the attack step of a technique you must prepare both your opponent's body and your own."* pag 36) [6]

We can then conclude that this is the phase during the throwing action when the thrower actually applies the throwing movement and constructs the throw. '*Tsuku*' means to prepare or build something and '*ri*' implies doing it to a person. The concept of *tsukuri* implies "preparing a person" to be thrown. What this all means is that the thrower moves his body into position to prepare his opponent to be thrown. *Kuzushi* and *tsukuri* blend together into a *kuzushi/tsukuri* movement rather than two separate movements. *Tsukuri* is dependent upon *kuzushi* to be effective.

### *2.3. Kake*

"*Kake*" translates into a phrase that implies "joining together." In other words, the attacker's body and the defender's body are joined or connected together at the instance of the attacker's throwing action with the attacker in control of how the bodies will move to the ground or mat. In fact, the *kake* phase is connected to the application of specific tools in order to carry out the throw. Which are these tools? Referring to Kanō's *Kōdōkan* classification for throwing an opponent there are only five tools: arms, legs, hips and one's own body mass (by dropping oneself either on the back or the side). *Kake* (application) implies various concomitant movements that make up the finishing part of technique towards obtaining. To accomplish this process may require involving global control as well as putting in place a barrier, or causing hyperextension to produce unbalancing one's adversary for the purpose of throwing. This phase of the throwing attack is the actual execution of the throw.

## *3. Classification of Throws*

The classification of standard jūdō throwing techniques (*Nage-waza*) as **Rational Sports Techniques** was born from the following didactic objective: **to group the standard techniques according to logical criteria for the purpose of facilitating understanding and of enabling useful systematic study**.

The first **Rational Sports Techniques** classification of *Kōdōkan*'s *Nage-waza* (1882) was very different from the one that is widely known today. In fact, all techniques were divided in three different groups. **Taosu koto** 倒す事 (Throw down techniques), **Otosu koto** 落とす事 (Drop techniques), and **Uchi tsukeru** 打付ける (Hit and strike techniques)[2]. However,

Kanō sensed that this kind of classification was too descriptive and not very useful. Consequently, the double problem of how to best organize these techniques and teach them was tackled and solved by Professor Kanō and his assistants in accordance with scientific considerations as far as understood in those days.

After some time (1885) Kanō implemented an innovative and refined classification based on proper proto-biomechanical methods. In this system the standard throwing techniques were classified according by "tools" or the parts of Tori's (the attacker) body which functioned as the major point of contact for energy transfer during the throw. Hence, this "*Kōdōkan Classification*" distinguishes the following groups of throwing techniques: **Te-waza** = shoulder, arm and hand techniques, **Koshi-waza** = hip techniques, **Ashi-waza** = leg techniques, **Sutemi-waza** = body-sacrificing techniques. This inventive simple, intelligible and almost perfect classification was rearranged four times (1885, 1922, 1982 & 2005). For these reasons the current *Kōdōkan* classification of throwing techniques has endured a long success despite several concerns, such as notably in reality it being unusual to see as tools: hands, hips or legs of *tori* working alone in throwing [5, 6, 7, 8, 9, 10].

## *4. Limitation of Kanō's Principles*

If we consider Kanō's systematization of throwing techniques one must affirm that it represents a for that time admirable expression of a fine application of rational and scientific thinking that is applied in a methodological way aiming to provide a pedagogical bedrock for Japanese jūdō wrestling techniques. However, a more cogent analysis shows its limitations, such as, for example, that it is in fact unusual to see hands, hips or legs of *tori* working in an isolated way as the tools carrying out a throwing technique. Furthermore, the *Kōdōkan* classification uses a different way to classify the body-sacrificing techniques, which are organized according to the side of the body of the thrower, and therefore, is not coherent with the parts of the body applying forces, which is the rationale underpinning the other three categories of the Kōdōkan throws classification.

Today, these historic times are long gone, and scientific knowledge is increasing very fast. Jūdō has become a worldwide-known Olympic sport, and the time is therefore appropriate to review and reassess Jūdō's systematic foundations in the light of modern biomechanical knowledge.

## *5. The Biomechanical Reassessment*

Biomechanics involves the precise description of human movement and the study of the causes of human movement [11]. The study of biomechanics is relevant to professional practice in many occupations. The coach who is teaching sport-specific movement techniques and the athletic trainer both use biomechanics to qualitatively or quantitatively analyze movements. The study of biomechanics requires not only an understanding of the structure of the musculoskeletal system and its mechanical properties, but also its deep interconnection with the central nervous systems, and with environmental circumstances and how they may affect the athlete [11]. Newtonian Mechanics is the branch of physics that measures the motion of objects, identifies forces and explains the causes of that motion.

Jūdō Biomechanics applies particularly to those aspects of human biomechanics that relate to this sport practice; it also includes the mechanical properties and design of sports equipment (*tatami*) and clothing (*jūdō-ji*). Knowledge of the mechanics of jūdō movements allows professional coaches to understand those movements, develop specific training exercises, and change movement techniques to improve performance, but also to invent new throwing techniques and to teach jūdō in a deeper and more useful way. The continuing growth of scientific knowledge presses for a substantive revision of principles well-established in the world of jūdō. The biomechanical reassessment of these jūdō foundations is conducted with the aim of offering the most lucid and systematic analysis, in this way helping both coaches and teachers in their daily work. In fact, a biomechanical reassessment will help not only the pedagogy of jūdō (clarifying and simplifying some didactic aspects), but also will facilitate a better understanding of other phenomena and help coaches to improve their athletes' performance by elaborating various competition-related aspects [9].

## *6. The Technical Principles of Jūdō Revised*

The revision of the foundations of jūdō we propose, employs the proper tools of Newtonian mechanics, and will focus on the core principles of jūdō teaching, which are represented by the triad: ***kuzushi, tsukuri, kake,*** as discussed previously. As part of this reassessment, we will expand the *kuzushi* (unbalance) concept as applicable to jūdō competition. In addition, we will offer a clear and easy classification and systematization of the *tsukuri* phase relying on what is known as **General Action Invariants** (GAI) [11,13]; this approach is very suitable for the purpose of teaching jūdō. Finally, a thorough revision of the *kake* phase is desirable, underlining what can be recognized as the two essential basic physical tools to throw down an opponent's body, *i.e.* a novel simplified biomechanical classification of throws based on the application of a "Couple" or a "Lever", and the **Specific Action Invariants** (SAI) [12.13]; this approach and classification are particularly useful to deal with newly created or different throwing techniques, which is an important advantage over the current jūdō pedagogical teaching and coaching framework.

### *6.1.    Kuzushi → Right unbalancing concept (breaking symmetry)*

All jūdō teachers and coaches are accustomed with the in jūdō universally known *kuzushi*, concept, as introduced by Professor Kanō. There is, however, a considerable conceptual distance between the theoretical explication (*Happō-no-kuzushi*) and its practical application in competition ! There are many reasons why it is not easy to apply the theoretical kuzushi concept under such circumstances, the most obvious one being the resistance produced by the adversary.

The theoretical explanation provided by Kanō was based on the physics of a rigid body. However, the human body is not a rigid but an articulated body. Modern biomechanics convincingly shows that the body's Center of Mass (COM) changes its position both inside and outside the body; by doing so, it changes the subject's stability situation also even without totally unbalancing the body.

In jūdō, we are used to consider balance while being in a neutral standing situation, that is, a position of unstable equilibrium, assuming that the COM of the athlete is still more or less under the navel in the origin of Athlete's reference system [11] connected to the body's well-known three planes of symmetry (frontal, sagittal and transverse). If the opponent's body is rigid, you can easily apply Kanō's unbalancing concept; if, however, the opponent's upper body part turns or bends to the side, you cannot.
In this last case, the COM shifts and changes position inside the body, and, consequently, both the body's stability and mobility are altered.

If one uses perfect timing (***hazumi*** 弾み, Litt.: 'recoil') then it is possible to advantageously use these transitory situations. This is the most useful and proper application of the unbalance concept in competition. If we consider that a human body is not a rigid system but an articulated one's, then we can understand the importance of a subtle application of the so-called "breaking of symmetry" to provoke unbalance. Such actions like bending or turning are easier to produce. They cause reduced mobility generated by shifting the COM inside the body and sometime outside, and in this way increase stability [11]. Normally, it is easier to provoke the opponent to bend or turn during grip fighting, because it is the opponent's natural reaction to being pushed, while he generally will have no conscious knowledge about the role of the breaking of symmetry concept.

"Breaking symmetry" changes the body's stability and slows down the opponent's capability to shift and compensate, in this way making the application of a successful throw easier.
This is what represents the first part of the "*advanced kuzushi concept*". The biomechanical explication splits the *advanced kuzushi concept* in two steps: first, before slowing down the opponent by breaking his symmetry by applying perfect timing (*hazumi*), and then following this step, a body collision (*ikioi* 勢い, Litt.: "with vigor & force") will occur which effectively ends the *kuzushi-tsukuri phase* and the *kake* phase commences. Normally, the body movements of both athletes will always increase the single unbalance situation that facilitates throwing action, but one still needs to be able to use this situation to one's advantage. The proper and effective use of such situation is often called *handō-no-kuzushi* (反動の崩し, Litt.: "reactional/recoiling unbalancing") by Japanese jūdō masters.

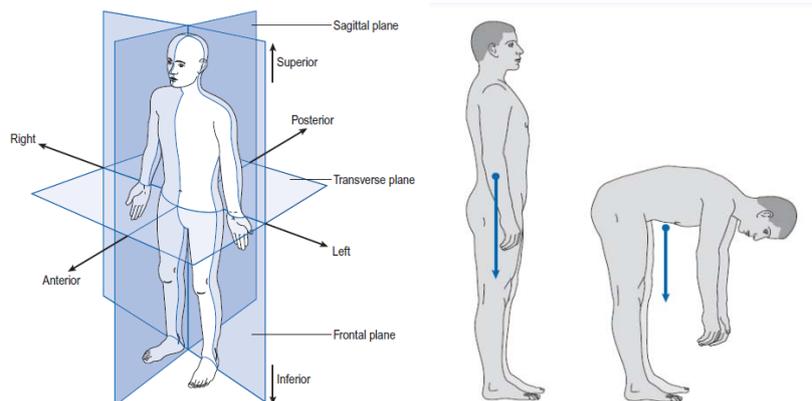

*Fig. 5,6: The human body's planes of symmetry. The center of mass (COM,) inside and outside the body.*

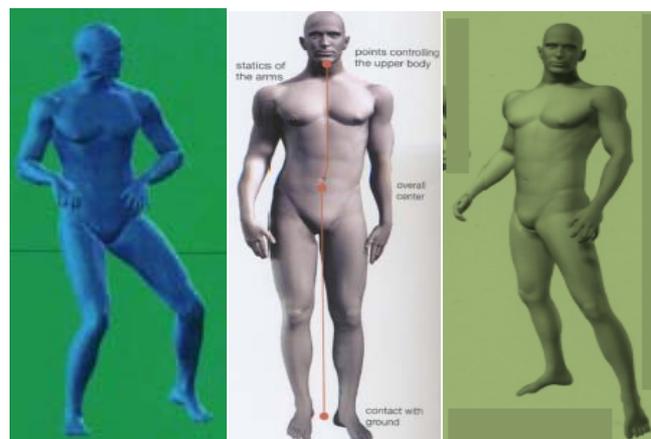

*Fig. 7: Three Broken Symmetry the COM inside the body leads to slowing down the opponent's capability of shifting that would enable him to recover his balance.*

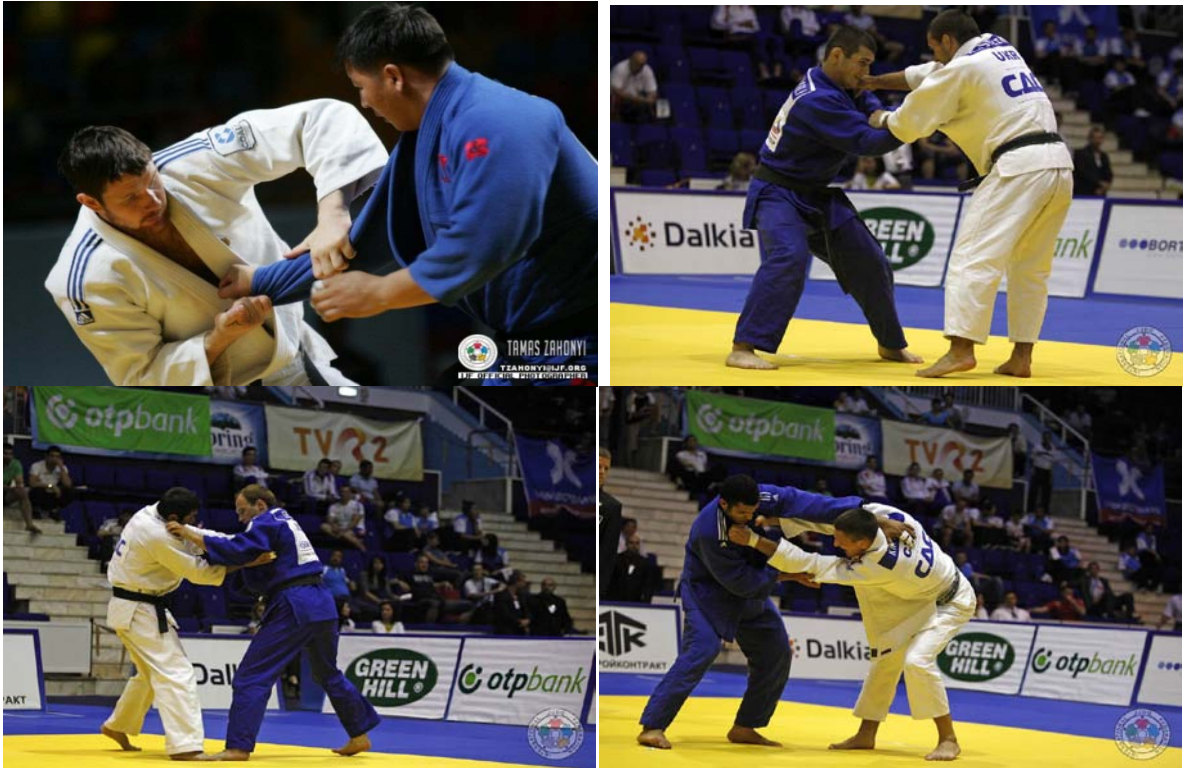

*Fig. 8-11: Examples of breaking symmetry during competition*

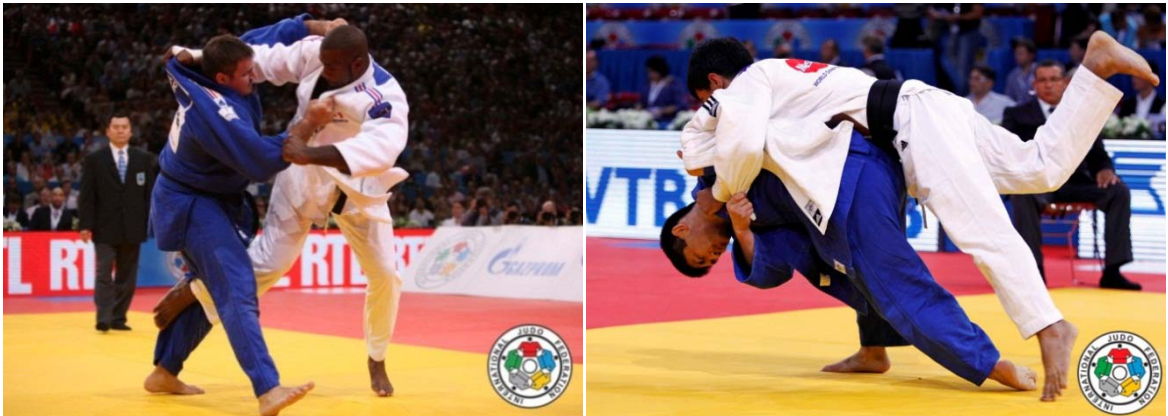

*Fig. 12-13: Examples of proper timing (hazumi) in Advanced Unbalance Concept*

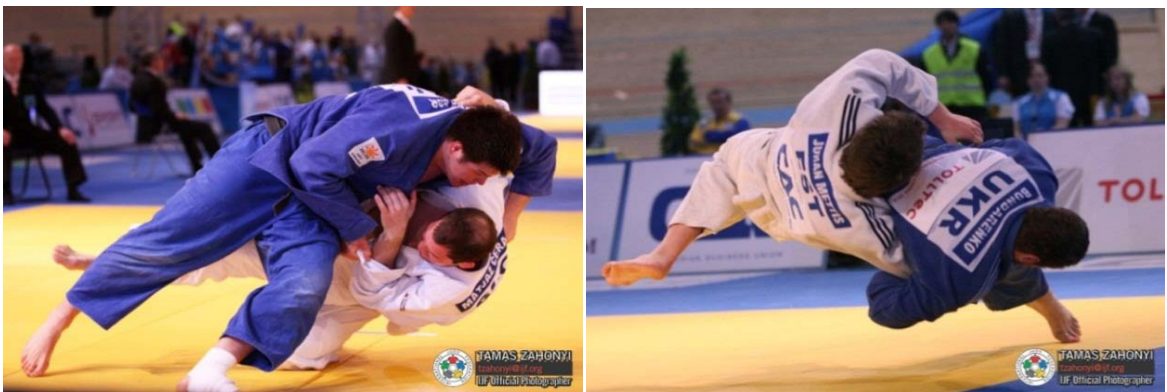

*Fig. 14-15: Examples of (ikioi) collision in Advanced Unbalance Concept*

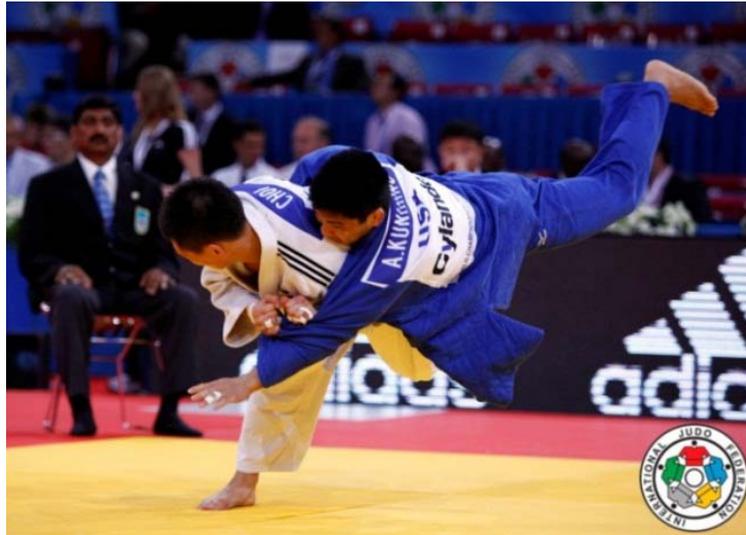

*Fig. 16: Example of simultaneous application of timing (hazumi) and collision (ikioi )*

It must be remembered that the effectiveness of breaking symmetry must be utilized at maximum by the ***Rotational Kuzushi*** application.

*Tai Sabaki* originates from higher or less (angular) velocity of execution and from optimal mobility of the hips compared to the axis of rotation and completes itself, at the end of ***Nage***, in a body's rotation at maximum of its flexibility

These movements, used also in Aikido, that in their basic form are named:

*Irimi* (入り身 Litt: is an entrance to the attacker approaching to his inside*;* to rotate entering into an opponent's attack)

and ***Tenkan*** (転換 Litt: is an entrance to the attacker that involves a step to the outside of his body; to rotate back and going off from the adversary).

They are enforced unlikely in their simplest form of pure rotation, but often in the real phases of a competition, the whole rotational ***Advanced Kuzushi Concept-Tsukuri*** becomes more complex, with others specific movements like: feet moving on the inside or outside of the attack direction; to and from (***Oikomi*** and ***Hikidashi***) to change suitably the relative distance; instantaneous rotations with a little jump (***Tobikomi****)* or with a feet pivot; cross-steps and other combinations with these movements utilizing always *Ikioi* and applying the final *Hazumi*. In such way breaking symmetry (as ***Dynamic Rotational Kuzushi***) becomes more effective and less energy wasting.

From a biomechanical point of view they can be classified as rotational movements with conservation of angular momentum or variation of angular momentum, exploiting the minor physiological capacity of the human body to resist to rotational stress.

The high performance Judo uses these rotational concepts as the most natural and appropriate to minimize the efforts and the energy, during the higher dynamic phases, in a modern competition as application of the ***Advanced Kuzushi Concept.***

Many people in jūdō do not know this advanced unbalance concept, and in fact err by breaking their own symmetry in this way aiding the adversary in his throwing action.

Hence, it is easy to understand that elite level jūdō athletes must practice a lot, focusing on how to break symmetry, *i.e.*: how to produce it, and how to use it in a useful way. This is a novel and original aspect of jūdō analysis full of new and interesting discoveries, which elite jūdō coaches must gain proficiency in to provide optimal feedback to their athletes

## 6.2. Tsukuri → General Action Invariants

*Tsukuri*, as previously stated, refers to the entire class of actions to bring the couple of athletes in the desired position in which one athlete can throw the opponent with minimal waste of energy.

Obviously, these movements are infinite in number; however, they pursue one common and definite objective: **to shorten the mutual distance.**

Indeed, this is a common aspect of the infinite number of situations that might arise [11,12]. Biomechanical analysis of this aspect shows some very interesting properties. It turns out that there are in fact only three classes of actions (trajectories of movements) that at the same time involve minimal energy and strive to achieve minimal distance.

In jūdō, that what we term *Action Invariant* refers to the minimal path, in time (like the Fermat principles in optics) of the body's shift, necessary to acquire the best *kuzushi* and *tsukuri* position for every jūdō throw.

Conversely, in those cases where it is actually possible to identify such a minimum action, or *Action Invariant* , the two following biomechanical axioms apply:

a) ***Best is the Judo Technique, minimum is the Athletes' energy consumption.***
b) ***Best is the Judo Technique, minimum is the Athletes' trajectory for positioning.***

As previously explained, we call these movement classes ***General Action Invariants (GAI).*** This term covers the whole range of body movements intended to both reduce the distance between both opponents and to optimize one's body relative to the adversary's body position.

Similar class of movements were found by S. Sterkowicz and coworkers in experimental study on competition: analyzing differences in the fighting methods adopted by competitors in All Polish Judo championship [13].

The biomechanical analysis of this class of invariant starts from one of the six ***Competition Invariants,*** Fig. 17

that are ***position of the couple of athletes carried out invariably in every competition, depending from the relative Kumi Kata of the couple of Athletes*** [8].

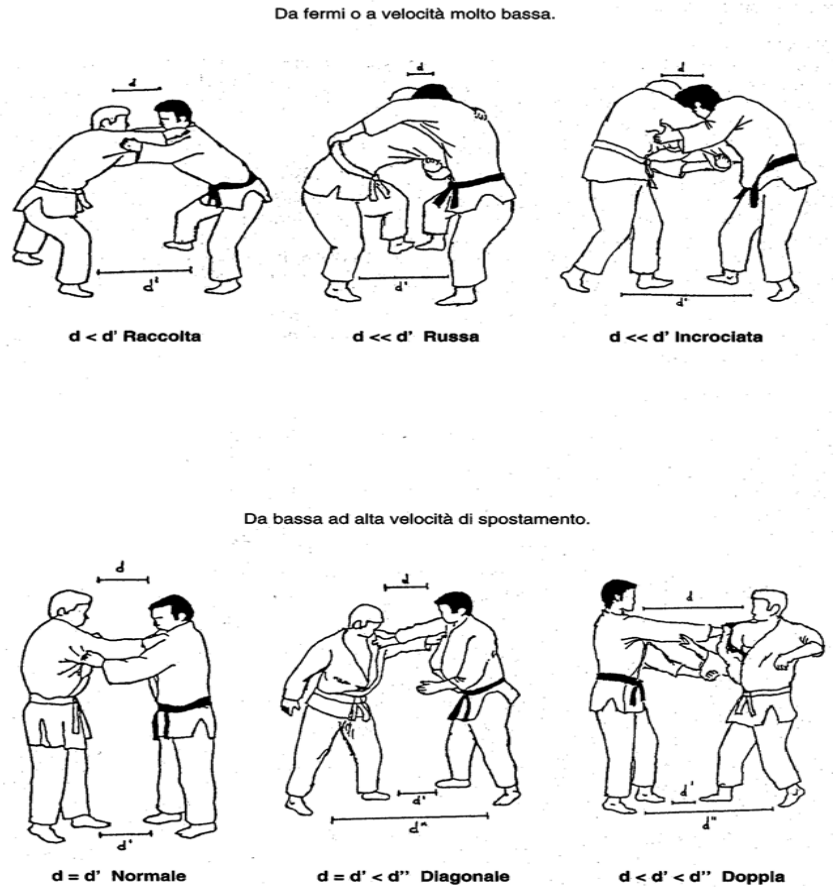

*Fig.17 Six Classes of Competition Invariants* **[8]**

This analysis specifies that only three *General Action Invariants* are included in the *kuzushi-tsukuri* phase for each *Competition Invariant*, namely:

1. Reducing the distance without rotation.
2. Reducing the distance involving complete (0° to 180°) rotation clockwise/counter-clockwise.*
3. Reducing the distance involving a half (0° to 90°) rotation clockwise/counter-clockwise.*

Each of these *General Action Invariants* is linked to the two biomechanics tools involved in effecting a throwing technique , application of: "Couple" or "Lever".

*These angles are valid for athletes in stationary position (study), but during a competitive match they may change if the athletes are moving; a classical example of this is the Japanese concept of Handō-no-kuzushi [Lit.: "reactional/recoiling unbalancing"]*

In terms of trajectory, the first class of **General Action Invariants** consists of nearly straight with specific direction; normally, the best straight inclination will be lateral, *i.e.*, in the direction of both adversaries' sides, because the human body is less skilled to resist unbalancing challenges into such directions. For example, within the throwing techniques group that consist of *Couple of Forces*-type of throws, this means that the couple of forces is applied laying in the frontal plane, such as is the case of *ō-soto-gari* or *okuri-ashi-barai*).

For the other two classes of **General Action Invariants** based on rotations, some interesting derivations can be made from the Poinsot geometrical description of a free-forces motion of a body. However the gross indication (which resembles circular trajectories) as available for our analysis from the Poinsot description, is quite acceptable as an 'indication' for real situations, even though jūdō movements have very complex three-dimensional trajectories that are very difficult in order to calculate their forces-time evolution in space.

## 6.3. Kake → from Five to Two Tools to Throw

The biomechanical analysis of jūdō throwing techniques in the sense of **Rational Sports Techniques** should be dealt with according to the following steps: firstly by simplification, and secondly, by generalization. By considering the three stages of a throwing technique as defined by Kanō as the core in his teaching of jūdō, we are able to simplify the different classes of forces involved in jūdō's throwing techniques. As previously indicated, these three steps are:

1. *Tsukuri* (preparatory movements aimed at throwing out of balance *uke*'s body)
2. *Kuzushi* (the final unbalancing action)
3. *Kake* (tools application aimed at throwing)

In a dynamic situation tsukuri starts simultaneously with the symmetry break that slows down the opponent and tsukuri and kuzushi overlaps each other till to the final unbalance action : collision of the two bodies or parts of them.
Subsequently, we can analyze the motion of *uke*'s body in space, omitting secondary and parasite forces. Then we generalize the classes of forces, putting out the inner physical principles of standard jūdō throwing techniques.
This method, when applied to *Nage-waza* (throwing techniques) allows grouping all (67 *Kōdōkan gokyō* and other) throwing techniques under two dynamic principles only [7, 8].
It is true that this is only one of many possible biomechanical classifications.
However, we prefer and select this classification because of its simplicity and immediateness. To identify the "**General Principles**", it is appropriate to first define the two corollaries relevant to the direction of forces (*i.e.*, Static Analysis), and subsequently to analyze *uke*'s body flight path (*i.e.*, Dynamic Analysis) and its symmetries.

*Static Analysis – Stationary Athletes*
These two corollaries determine the whole directional problem when applying static use of forces to execute jūdō throws.

*Unbalance*
*1. Forces are effective and can be applied, in the horizontal plane, at each angle (360°).*

Within this category we consider the biomechanical problems that pertain to the forces employed for unbalancing *uke*'s body (*i.e.*, the *kuzushi-tsukuri* steps in **Kanō's definition**).

*Throws*
2. *Forces are effective and can be applied in the vertical plane, over a range that corresponds to almost the entire width of a right angle (90°).*
Within this category we find the biomechanical problems that pertain to the forces employed for throwing *uke*'s body. (*i.e.*, the *kake* steps in **Kanō's definition**).
The real limits of forces that are active in throwing can be reached at an angle of nearly 45 degrees up or down a horizontal line. Beyond these angles, the resistance produced by *uke*'s body structure or by the force of gravity may still allow throwing, albeit with far more waste of energy or by using one's own body mass to assist in completing the throw.

*Dynamic Analysis – Moving Athletes*
These corollaries determine the whole directional problem of dynamic use of forces to execute throws in motion [7-9].

*Advanced Dynamic Unbalance*
1. *Breaking the body's symmetry will produce a shift of the COM into the opponent's body, thus increasing his stability and impairing his shifting capability*
2. *Timing is used to fit the two bodies' unbalanced positions, and Unbalance is concretized when these bodies collide (two-bodies collision)*
3. *Forces are always effective and can be applied at each angle over a range of 360° within the horizontal plane.*

Unified under these terms are the biomechanical problems of forces employed for unbalancing *uke*'s body during competition (*Tsukuri/kuzushi* steps in **Dynamical Unbalance Situation Throws**).

4. *Forces are effective and can be applied, on the vertical plane, nearly for the width of a right angle (90°).*

Within this category we find the biomechanical problems that pertain to the forces employed for throwing *uke*'s body by the two biomechanical tools "Couple" or "Lever" (*i.e.*, the *kake* steps in **biomechanical analysis**). The real limits of forces that are active in throwing can be reached at an angle of nearly 45 degrees up or down a horizontal line.
Beyond these angles, the resistance produced by *uke*'s body structure or by the force of gravity may still allow throwing albeit with far more waste of energy or by using one's own body mass to assist in completing the throw.

*Principium of composition of forces – Study of flight paths and their symmetries*
If in space the composition of forces at the same time adheres to the previous corollaries, then the solution of dynamic problems (considering time) involves the study of the flight paths and their symmetries in space [7-9]. These trajectories along which *uke*'s body moves as a result of the throw that is performed on him, consists of two options (types) defined according to the choice of tool to throw:

1. *Circular paths*
2. *Helicoidally paths*

These two trajectories represent the geodetic expression of specific symmetries (Spherical *vs.* Cylindrical). This implies again that it is the **Least Action Principle**, which applies to the throwing trajectory. If we think of the corollaries for Static and Dynamic Unbalance, of the direction of forces, and of the *uke*'s trajectory, one can define the **two tools** showing the inner mechanisms of the chosen throwing techniques used on the *uke*. [7-9].

*A) First Tool:*
*Techniques where tori makes use of a "Couple of Forces" for throwing uke.*

*B) Second Tool:*
*Techniques where tori makes use of a "Physical Lever" for throwing uke.*

Consequently, we can also assess the basic conceptual and biomechanical steps (useful for coaching and teaching) of jūdō attacks that occur during Dynamic Competition.
These steps really reflect a continuous fluent movement:

1. **First: breaking the Symmetry to slow down the opponent** *(i.e., starting the unbalancing action)*
2. **Second: timing,** *i.e.*, **applying the "*General Action Invariant*", with simultaneously overcoming the opponent's defensive grips resistance,**
3. **Third: a sharp collision of bodies** *(i.e., the end of unbalance action)*
4. **Fourth:**
    A. **Application of "Couple of Forces" tool** (*i.e.*, **the type of throw), without any need of further unbalance action,**
        Or
    B. **Use of the appropriate "Specific Action Invariants", needing to increase unbalance action, stopping the adversary for a while, to apply the "Lever" tool** (*i.e.*, **the type of throw) in Classic or Innovative or New (Chaotic) way.**

The previous steps represent that what are the simplest movements, which occur in a connected way and which are usually used to throw the opponent. Very often though, far more complex situations can arise under real fighting conditions. These complex situations which have evolved from the simple steps explained above depend on the combination of attack and defensive skills of both athletes. However, the actual *Collision* step is very important for applying any real throwing technique. This kind of collision is almost a rigid body collision [23-26].Nevertheless the use of rotational forces approach in ***Advanced Kuzushi Concept*** seems the most effective one's in term of score. Fig 18

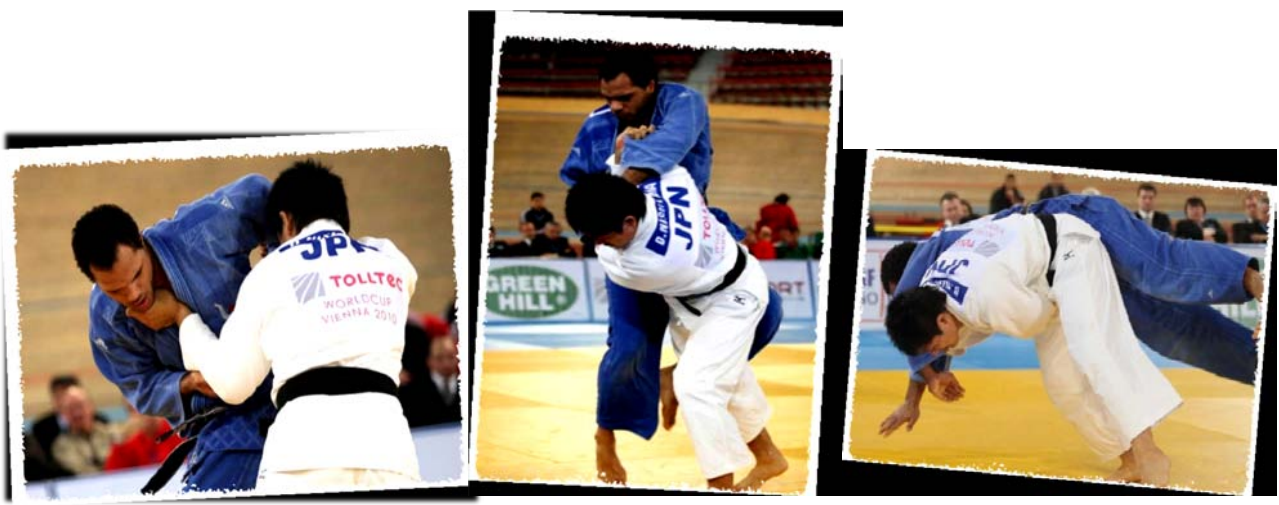

*Fig.18 Application of effective Advanced Kuzushi Concept in High Performance: Breaking Symmetry then [GAI+ Hazumi +SSAI]  + final Ikioi in a rotational (sided forces application)*

*Couple of Forces throws*
In a throwing action, the collision is the main movement of inferior dynamic chains in the Couple groups (arms and leg, trunk and leg). The collision as applied in these techniques is an "off-center type of collision" with friction. The **off-center collision** does not normally follow the connecting line of the athlete's center of gravity; hence there is a torque present. The *uke*'s body begins to rotate, in this movement being also assisted by the other side of the couple application. Because the athletes' bodies represent macroscopic bodies we can use the goal of **penalty models** to study their collision action. ( See Physics and Mathematical framework paragraph 8).

*Physical lever throws*
Normally, for the Lever group of throwing techniques, the actual collision can occur at a stopping point. However, in real contest situations before the stopping point is applied, the Dynamic Unbalance Action ends with a collision between *tori*'s and *uke*'s bodies. That happens, very often, in the Physical lever throws: group of "minimum arm", because in such throws opponent's bodies join.
Important parameter in high-level jūdō competition is the athlete's body mass (weight).
What we have explained above depends on fast-shifting mobility and high-speed applications. However, these capabilities are very often inversely proportional to one's body weight. The actual fighting style changes with body weight, and the general lack in fast-shifting capability —the well-known slow pace of heavyweights— produces an increase in application of forces. This in itself provides a justification to investigate the proper use of the "breaking symmetry tool" during high-level jūdō contests. This kind of knowledge helps to drive and simplify the throwing phase according to a more subtle understanding of the classic Kanō unbalancing concept.

## 6.4. *An expanded concept of jūdō throws*

Biomechanical analysis allows us to single out the only two tools utilized to effect jūdō throws. These tools: "Couple" and "Lever" are the basic way to apply forces on the opponent's body in order to throw this opponent in all circumstances. When these tools are well understood, every athlete can apply them in an infinite way. This is an expanded concept of jūdō throws, since it is definitely not connected to the 40 rational throwing techniques that form the basis of the *Gokyo* as defined by Professor Kanō. One of today's problems in jūdō is the introduction of new techniques on jūdō mats worldwide. Many of these techniques cannot be categorized according to Kanō's five principles, ( hand, leg, hip, and Side of body falling) because these principles are not "well defined", and too connected to the 40 techniques formalized in 1920 by Kanō and his assistants. Despite these limitations, efforts by Toshirō Daigo (former Chief-Instructor of the *Kōdōkan*) and the *Kōdōkan*'s Technical Commission, have resulted in extending the number of jūdō throws from the previous 40 to 67 throws with more or less 35 variations, all organized according to Kanō's five principles [7].

## 6.5. *Specific action invariants*

Analyzing the judo throwing techniques from the biomechanical point of view, paying attention to the role played by kinetic chains (arms and legs) some interesting remarks are found.
For the "Couple" group the role played is almost "monotonous" (the application of force parallel into the "Couple" produced generally by system leg/s - arm/s or two arms), for the techniques of lever system throws, biomechanics allows us to identify the specific, more complex, role played by both superior and inferior kinetic chains.

In respect of the definitions previously discussed and utilized for the *General Action Invariants,* we can call these specific movements or actions "**Specific Action Invariants** (**SAI**)", which can be divided into **Superior Specific Action Invariants** (**SSAI**) and **Inferior Specific Action Invariants (ISAI)** [12, 14, 18]. Biomechanics then is able to also explain how these Invariants principles work during the process of applying jūdō throwing techniques.

### *6.5.1. How SAI work in the Lever-based Throwing Techniques Group*

In general, considering the degrees of freedom involved in the superior kinetic chains related to body movement, it is easily understood that all the potential movements in the *Specific Action Invariants,* for the superior chain (*SSAI*) in the *kuzushi* actions, are connected to the three degrees of freedom of the acromion joint (shoulder).

For the (*ISAI*), the most important part of the inferior actions control two joint systems, namely, hip and knee, and to a lesser extent the foot/ankle; the first joint system (hip & knee) is mainly used for adjusting one's body in an optimal position in relation to *uke*'s body, and the second joint system (foot & ankle) is utilized for applying the fulcrum as part of the lever action). It is also important to note that both *Specific Action Invariants* must be connected to each other in one harmonic way to produce a proper and effective throw as a result. The *kake* phase, for the physical lever-type of techniques, is the result of a well-coordinated and -interconnected action performed by both kinetic chains in different time sequences. First, there is the superior-chain open space that involves the body as part of the opponent's grip; secondly, there is the *General Action* (reducing the distance) that is pursued and harmonically followed up by the coordinated and connected work of both *Inferior* and *Superior Action Invariants* as achieved through the abdominal and trunk muscles.

These techniques need more skill in harmonic chains-connected movements, than Couple of Forces-type of techniques; in fact, often such techniques are ineffective because of a lack in harmony in one of the preceding movements halts the throw, essentially preventing any score or result.

### *6.5.2. Work of GAI and SAI in classic Japanese jūdō throws*

In terms of teaching pedagogy it is useful to explore how and in what way GAI and SAI are connected to classic Japanese throwing techniques. For example, the same *General Action Invariant* (**full rotation**) which (considering only the inferior chains) starting at the top and moving downwards splits up into three *Inferior chain action invariants* conform to three well-known classic jūdō techniques, namely (standing) *seoi-nage*, (kneeling) *seoi-otoshi* and (seated) *suwari-seoi*. All these *kuzushi/tsukuri* movements flow over into the *kake* phase of a physical lever-type throwing application with variable arm, from the most energy-wasting variant (standing) till to the lesser energy-wasting variants (seated) (See fig. 19, 20).

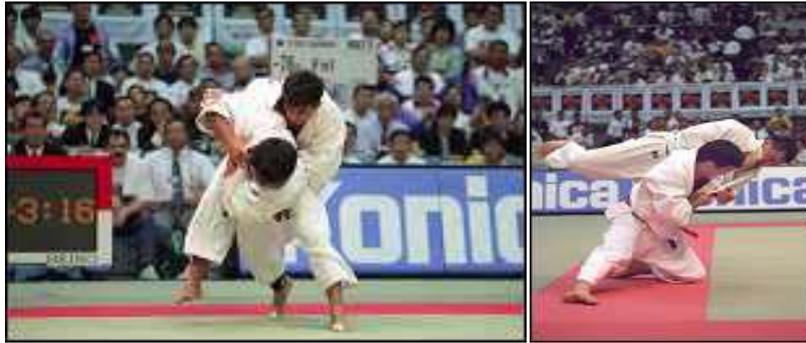

*Fig. 19, 20: Inferior Chain Action Invariant (Seoi); Inferior Chain Action Invariant (Seoi-otoshi) (Finch).*

The same *General Action Invariant* (**direct reduction of distance without rotation**) which (considering only the inferior chains) starting at the bottom and moving upwards splits up into three *Inferior chain action invariants* also conform to three well-known classic jūdō techniques, namely, *sasae-tsuri-komi-ashi*, *hiza-guruma*, and *tomoe-nage*. If one looks at the inferior kinetic chain, it is only the leg and foot of *tori* which move up from *uke*'s ankle, to his knee, and finally, his abdomen. All these *kuzushi/tsukuri* movements flow over into the *kake* phase of a physical lever-type throwing application with variable arm, from the least energy-wasting variant to the most energy-wasting variants.

The same *General Action Invariant* (**full rotation**) which (considering only the inferior chains) starting at the bottom and moving upwards splits up into three *Inferior chain action invariants* conform to three well known classic jūdō techniques, namely *tai-otoshi*, *ashi-guruma,* and *ō-guruma*. If one looks at the inferior kinetic chain, it is only the leg and foot of *tori* which move up from *uke*'s ankle, to his knee, and finally his abdomen. All these *tsukuri* movements flow over into the *kake* phase of a physical lever-type throwing application with variable arm, from the least energy-wasting variant to the most energy-wasting variants. The Fig.21 shows the inferior-chain action invariants connected to *sode-tsuri-komi-goshi*.

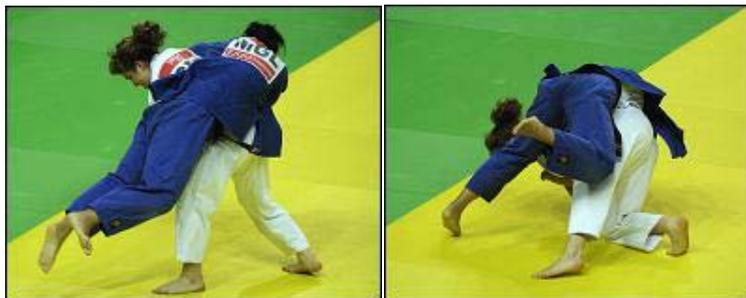

**Fig. 21 : Inferior Specific Action Invariant for an "innovative"** *Sode-tsuri-komi-goshi henka* **(variant) (Finch)**

Among the lever group of jūdō throwing techniques there exists a "Maximum-harm subgroup" in which the **Superior-chain Action Invariants** play the most important role; these include techniques such as: *uki-otoshi* and *sumi-otoshi*, which are characterized by the absence of any body contact, and where the fulcrum lies under *uke*'s and consists of the friction between *tatami* and foot. Now we can understand the main use of the **Superior chain Action Invariants**: generally, this class is employed when force application becomes involved. In this way it is possible to define the main useful angles for *kuzushi* required and essential for carrying out these kinds of techniques, because obviously it is not mechanically possible to apply a physical lever principle without the application of a force distant from the fulcrum. For this reason, as wells as because of the required coordination and connected movements, these type of techniques are biomechanically rather complex. To be applied in a useful way in a jūdō contest situation, this type of physical principle, irrespective of the shifting velocity of the Couple of Athletes, requires a minimal stopping time.

### 6.6. How Classic, Innovative and New (or Chaotic) throws are built.

Before defining the concepts "Innovative Form" and "New or Chaotic Form" jūdō throwing techniques, it is important to consider that many countries have made and are still making new technical contributions to *Kōdōkan* jūdō. The motivation towards victory and the ability to overcome opponents represent the root of this evolution (without implying either a positive or pejorative meaning, and merely using the term in a sense of "changes over time"). Some examples of these newly added techniques entries are: the **Korean seoi-nage**, the **Cuban – sode-tsuri-komi-goshi**, the **Korean sukui-nage**, the **Russian gyaku-uchi-mata** by Shota Chochosvili or the **Khabarelli**, which was a well-known "free wrestling" techniques [9] or the **Armbar** by Neil Adams, to name just a few. These techniques cannot be found in most standard jūdō books. The main analyst of the Innovative aspect of these throws is **Roy Inman** from England [21-24]. These kinds of new throwing technique variations very often arise from the will to apply classic techniques in highly changeable dynamic situations. In doing so, their users may be able to catch their opponent off-guard with such high dynamic throws. The following figures show a few examples of such Innovative throws from Russia and Cuba

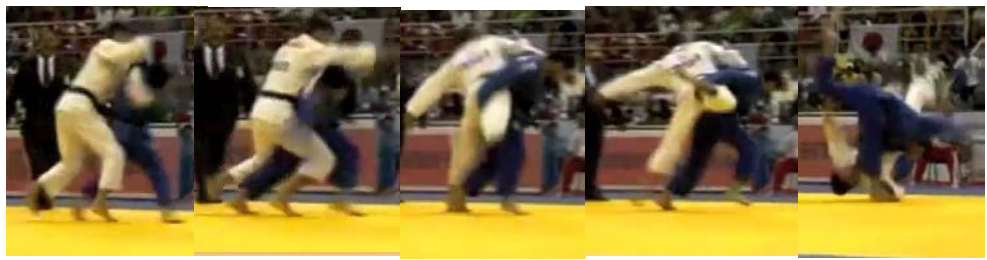

*Fig. 22 : Cuban sode-tsuri-komi-goshi with one-sleeve grip.*

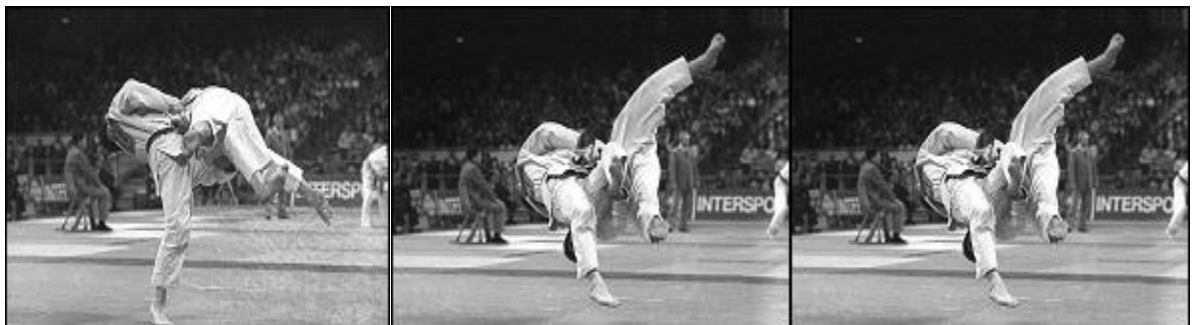

*Fig. 23: Russian gyaku- (or ushiro-) uchi-mata by Shota Chochosvili (Finch)*

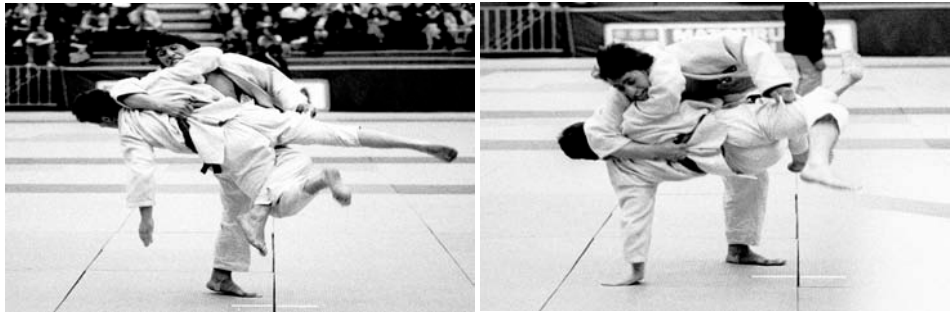
*Fig.24: Khabarelli-type Throws (Finch)*

These situations illustrate that there appears to arise a problem regarding the name of these jūdō throws. From there the question arises whether it is necessary to use a new name for different throws ?

Before addressing this question it is appropriate to consider the words of the late Kazuō Kudō, one of the last surviving students of Jigorō Kanō, about the names of classic jūdō throws:

"*Jūdō names fall into the following categories:*

1. *Name that describe the action:* ***ō-soto-gari, de-ashi-barai, ō-uchi-gari-gaeshi.***
2. *Names that employ the name of the part of body used:* ***hiza-guruma, uchi-mata.***
3. *Names that indicate the direction in which you throw your opponent:* ***yoko-otoshi, sumi-otoshi.***
4. *Names that describe the shape the action takes:* ***tomoe-nage*** *('tomoe' is a comma-shaped symbol).*
5. *Names that describe feeling of the techniques:* ***yama-arashi, tani-otoshi.***

*Most frequently, jūdō techniques names will use the content of one or two of these categories.*" (From: Kudō Kazuzō: *Dynamic Judo Throwing technique* [6]. Appendix p. 220).

As *jūdōka*, we are accustomed to using the same name for a given jūdō movement everywhere in the world. This standardization of jūdō technique names is very useful as it allows understanding each other beyond any language barrier.

Biomechanics, however, does not apply to names but to basic mechanical concepts. We opine that for reasons of optimal communication it is preferable to not introduce new names for jūdō throws. However from a biomechanical point of view, it is important to define and understand how such techniques may be developed. Therefore, one can then in terms of biomechanical analysis conclude that:

***"Innovative Throws" are all throwing techniques that keep alive the formal aspect of classic Jūdō throws, and differ in terms of grips and direction of applied forces only.***

In our definition, **Innovative Throws** are **henka** (variations) applications of classical *Kōdōkan* throwing techniques, which biomechanically are either Couple of Force-type or Lever-type techniques, while it remains easy to still recognize a basic traditional technique (40 *gokyō* throwing techniques) in them.

However, there are other "non-rational or non classic" solutions applied in competition and which are different form 'Innovative' (*henka* Throws), which we define as "**New or Chaotic Forms**". Oftentimes these **New or Chaotic Throws** are mainly limited by the class of lever group. When one analyses these types of throws more in depth, then the real difference between the goals of kuzushi/tsukuri in both biomechanical groups of throws will become clear.

In fact, the final *kake* phase of Couple of Forces-type of techniques does not require any further refining movement.

The Physical Lever Group-type of techniques, however, requires the addition of some specific refining movements depending on the proper positioning action of both kinetic chains. This is the major difference between the different techniques that continue into a *kake* of execution phase by Couple -type throws *vs.* Lever-type of throwing techniques.

All throwing techniques which rely on a Physical Lever must be refined by means of so-called ***Specific Action Invariants (SAI)*** that are common to each class of these throws, and that are related to both the effective positioning of the kinetics chains and to how to apply the necessary forces to unbalance the *uke* before throwing. The ***Specific Action Invariants*** (**SAI**) are infinite in number and are related to the superior and inferior kinetic chains motion and proper positioning necessary for the *tsukuri* and *kuzushi* phases. This analysis leads to some further interesting considerations.

For example, among throwing techniques, Couple -type of throwing techniques are independent from kuzushi; unbalancing could help the throwing action, but is not necessary.

Conversely, Lever-type throws need all two ***Action Invariants*** to be performed successfully; as a complex motor skill, they are more difficult to perform.

This evaluation shows that it is easy to understand that innovations are possible both in the Lever and Couple groups of techniques, because in both groups it is possible to apply a wide variation of grips and hence producing forces in different yet useful directions.

Consequently, both in the Lever-type and Couple -type Groups, ***innovation*** (*henka*) is connected to different directions of ***Superior Specific Action invariants (SSAI)***, but in the Lever-type throws group, innovation could also arise from a different collocation of ***Inferior Specific Action Invariants*** (***ISAI***). Because Couple of forces-type throws are technically easier throwing movements (***GAI +Couple+ Kake***) Innovations in this group are possible, based on *tori*'s stereotyped body movement which is responsible for the *kake* action; therefore, very few ***New or Chaotic*** are created within this group. Conversely, in the more complex built Lever-type of throwing techniques ***[GAI + (SSAI + ISAI) + Lever + Kake]*** the options to establish ***New or Chaotic Forms*** that are based on simple or complex variations of the (***SSAI*** or ***ISAI*** or both) are easier to realize.

*"**New or Chaotic Throws" principally belong to the Lever-type Group, and are characterized by the application of different grips positions (SSAI) which applying force in different (nontraditional) directions while simultaneously applying (ISAI) in non-classical positions.***

The feeling of 'different' is produced by new or chaotic non-identifiable throwing techniques from a view point of those accustomed to seeing classic *Kōdōkan gokyō* expressions of throws. Throwing techniques produced in such way are normally called 'new', and they present a number of difficulties in facing them. The following figures show some of these *Innovative Techniques* (some of them studied by Roy Inman), which belong to either the Couple -type or the Lever-type group of throwing techniques; they are characterized by variations to the stereotypically body movement by connecting different kinds of grips that allow only small directional variations (*Innovation*).

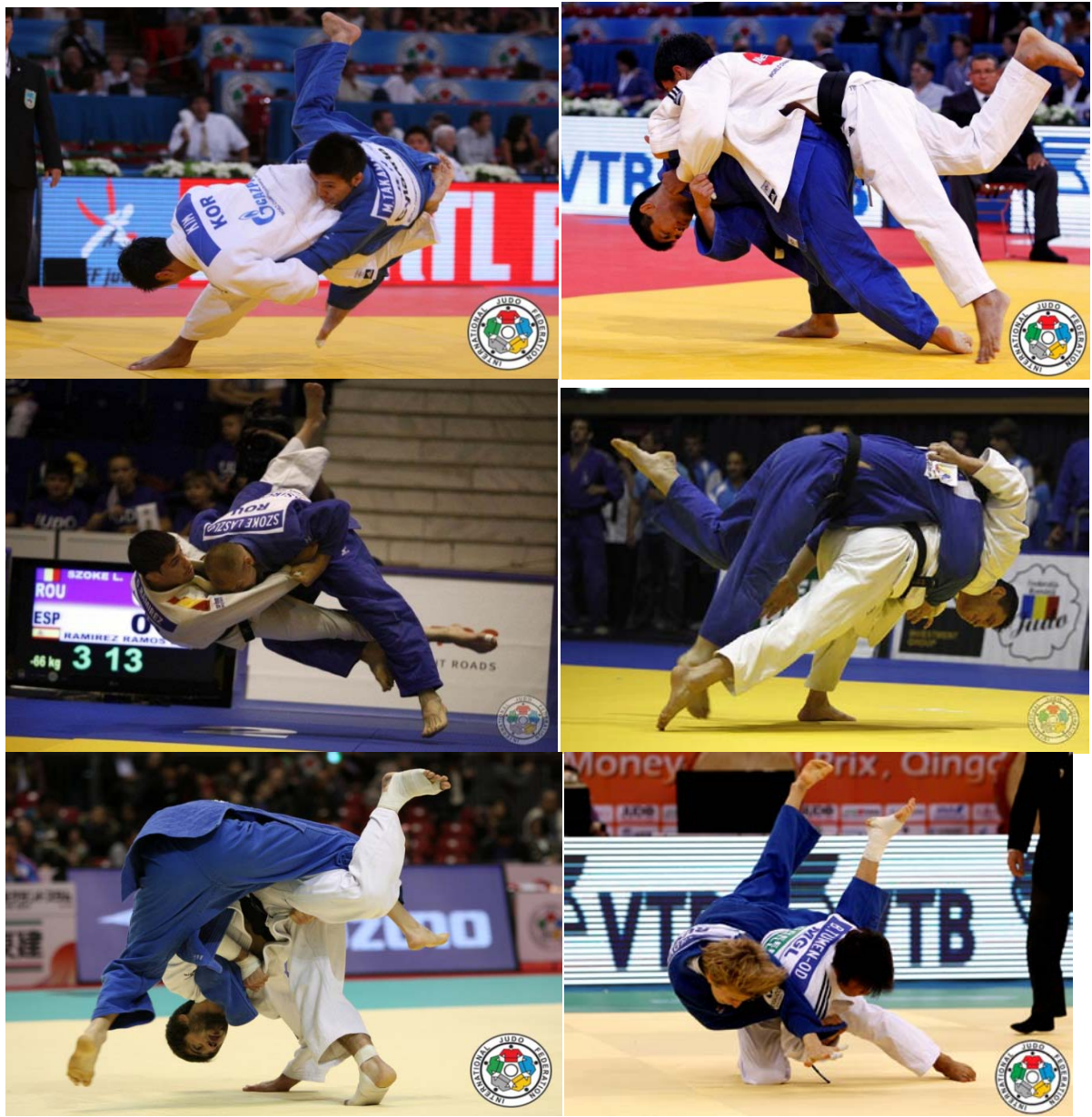

*Fig. 25-30: Application of Innovative Forms of throwing techniques; all these throws represent henka of Kanō's Classical Throwing Techniques (different contact points, grips, and direction of forces)*

At the end there are shown some "*Chaotic Form*", in which is easy to see the original use of both *SSAI* and *ISAI*

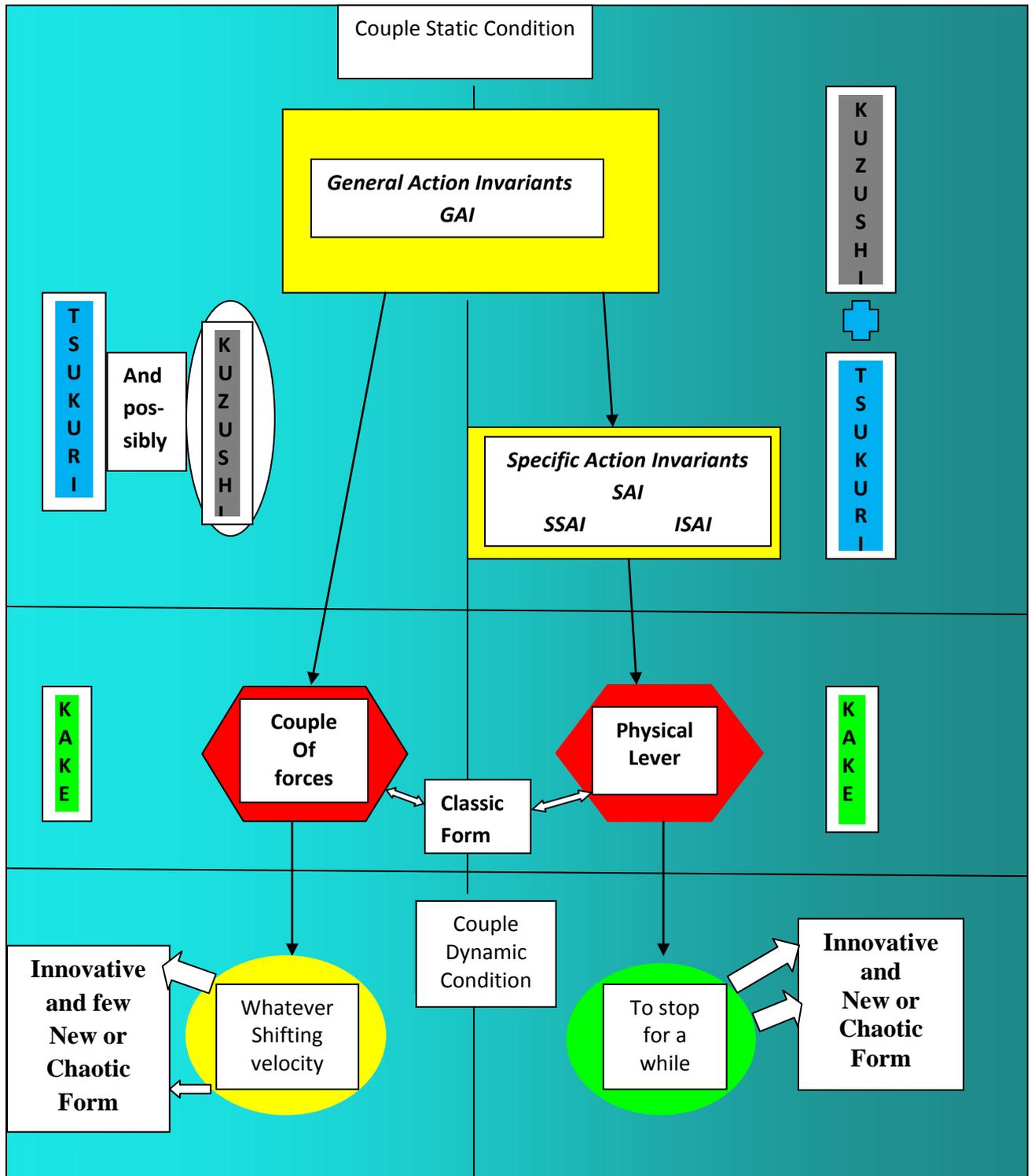

*Diagram 1: Summary of the Kuzushi Tsukuri Action Invariants connected to Kake phase and Classic or Innovative and New (or Chaotic) Form of throwing techniques*

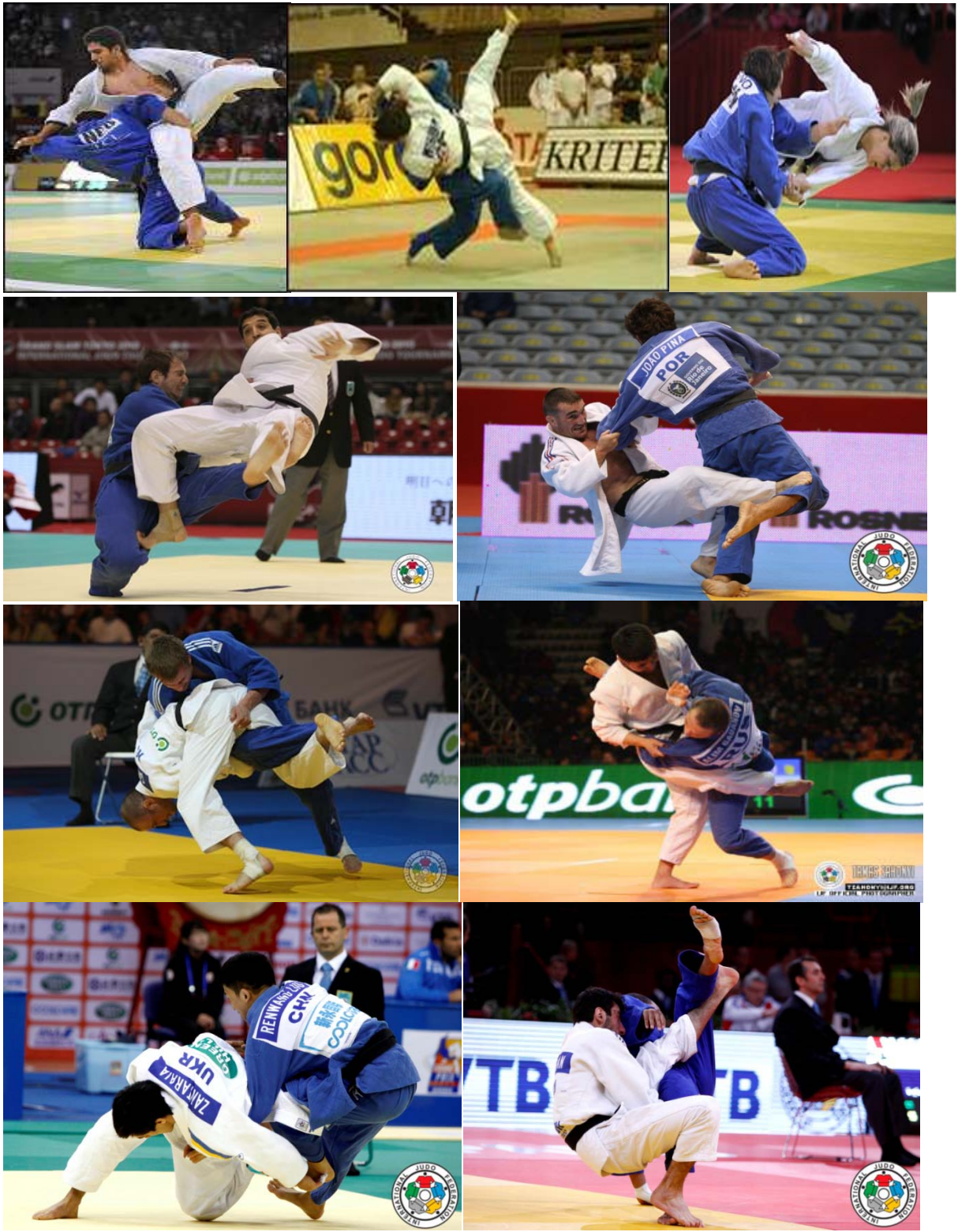

*Fig. 31-39: New (or Chaotic) Forms of Throwing; all these throws represent applications of the two biomechanical principles but in a form very different from Kanō's classification.*

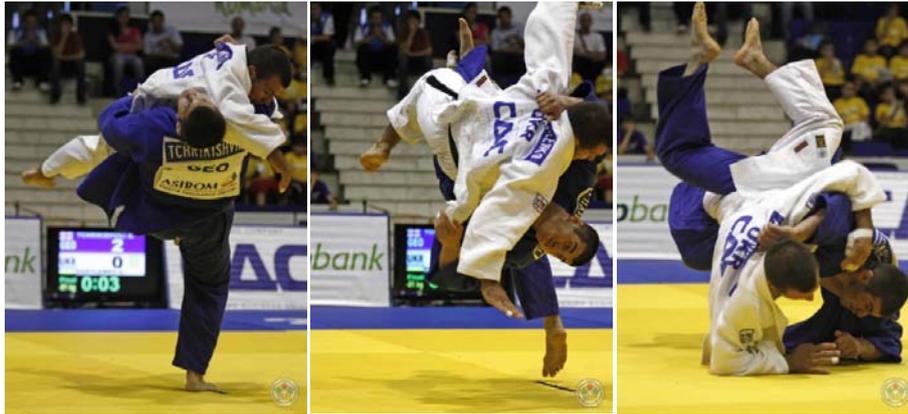

*Fig. 40: Gyaku- or ushiro-uchi-mata as an example of Innovative (henka) or New (Chaotic) Form of throwing ?*

## 6.7. Biomechanical Classification of Jūdō Throws

In the light of our analysis we believe that clarification of the basic physical principles and the evidence that supports basic technical actions allows classifying all possible jūdō throwing techniques according to clear scientific principles and rationale [7-10, 27]. In term of Biomechanics all throws could be classified into only two groups:"Couple Techniques" and "Lever techniques".

### A) Techniques produced by a Couple as tool.

In this first group we find all throws that are produced by sweeping away legs and simultaneously pulling or pushing *uke*'s body in the opposite direction. The techniques within this "Couple of forces group" can be classified according to the parts of *tori*'s body which apply and realize the couple of forces on *uke*'s body; these parts are, either: two arms, an arm and a leg, the trunk and a leg, the trunk and arms, or two legs. This biomechanical approach and classification is able to demonstrate a likeness between these techniques that is not evident in standard technique classifications. For example, the front-to-back asymmetry of the human body clarifies the astonishing biomechanical alikeness of several techniques: *ō-soto-gari*, *uchi-mata* and *harai-goshi* represent the same stereotyped movement for *tori* in terms of applying the couple of forces on *uke*'s body [8,9] This is clearly illustrated in the next three sequences showing Kōsei Inoue demonstrating at Bath University in the UK.

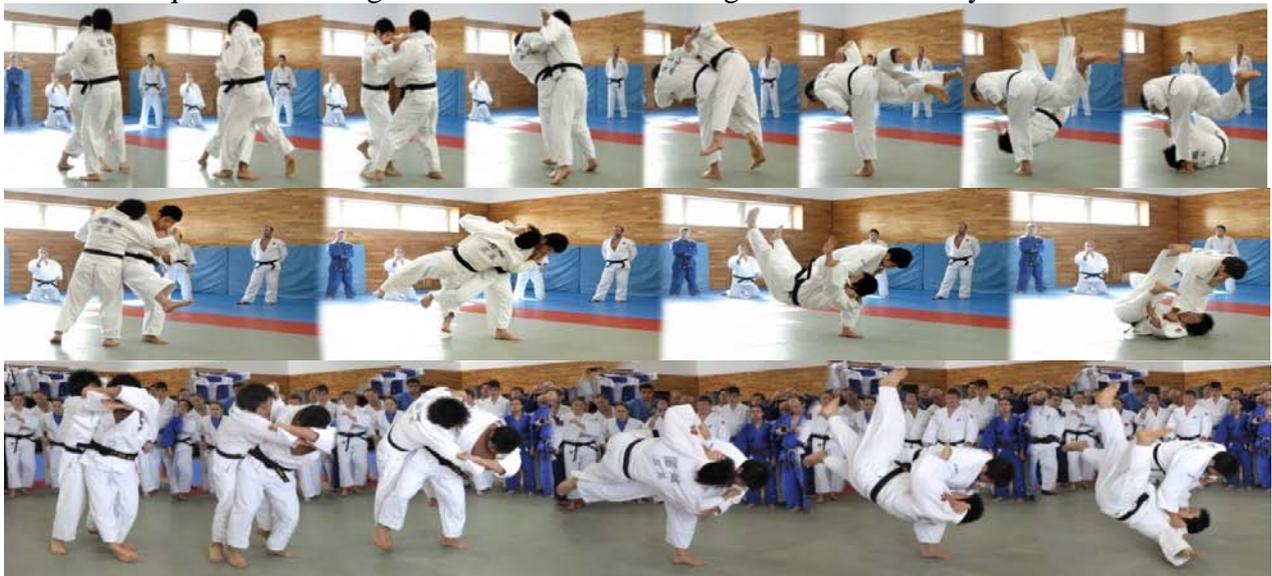

*Fig. 41-43: Similarities in tori's (stereotyped) arm / leg action during execution of Couple type techniques, namely three different classics: O Soto Gari - Uchi Mata - Harai Goshi*

It is very interesting to note that most throws within the couple-of-forces (the ones applied by *tori* standing on one leg) group of throwing techniques can be reduced to a single basic action of *tori:* rotation on the coxo-femoral articulation with three degrees of freedom, each of them conforming to one of the human body's three planes of symmetry[8,9,10].

First: rotation of the trunk-leg set on the coxo-femoral articulation around a horizontal lateral-lateral axis of rotation (= motion occurring within the sagittal plane). Fig.44

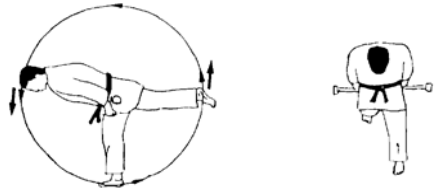

*Fig. 44: Motion within the Sagittal plane*

Second: rotation of the trunk-leg set on the coxo-femoral articulation around a horizontal lateral-lateral axis of rotation (= motion occurring within the sagittal plane) but in the inverse direction. Fig.45

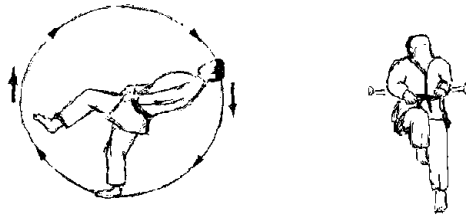

*Fig. 45: Inverse Motion within the Sagittal plane*

Third: rotation of the trunk-leg set on the coxo-femoral articulation around a horizontal antero-posterior axis of rotation (= motion occurring within the frontal plane).Fig. 46

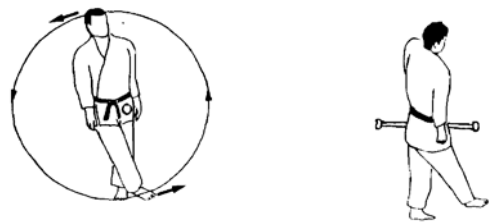

*Fig. 46: Motion within the frontal plane*

Fourth: rotation of the trunk-leg set on the coxo-femoral articulation around a vertical axis of rotation (= motion occurring within the transverse plane). Fig.47

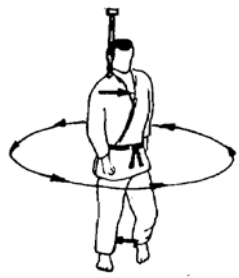

*Fig. 47: Motion within the transversal plane*

These examples illustrate the fundamental role played in this group of throwing techniques by ***COXO-FEMORAL ARTICULATION***, and entail that the athlete choosing to gain proficiency in these techniques and to use them with efficiency should be able to involve this articulation with a great flexibility.[8, 9,10]

*B) Techniques produced by a physical lever as tool*
In the second group we find all throws that are produced by turning *uke*'s body around a stopping point (hip, leg, foot, *etc.*). The techniques of this "Lever Group Techniques" can be classified by length of the lever applied on *uke*'s body, namely: minimum lever length (fulcrum under *uke*'s waist), medium lever length (fulcrum under *uke*'s knees), maximum lever length (fulcrum under *uke*'s malleoli), and variable lever length (variable fulcrum from the waist down to *uke*'s knees). Within this group, throws employing "minimum arm" as a lever, are, however, energetically unfavorable (*i.e.*, require the greatest force), which is why during contest situations jūdō athletes tend to prefer turning these into throws of variable lever length hence bringing down the fulcrum under *uke*'s waist more and more, which implies less waste of energy. [8, 9,10] In the Following figures are described the symmetries of Lever throws and the "geodetics" of uke's body during flights . ( Helicoidally for Lever Circular for Sutemi throws). Fig 48

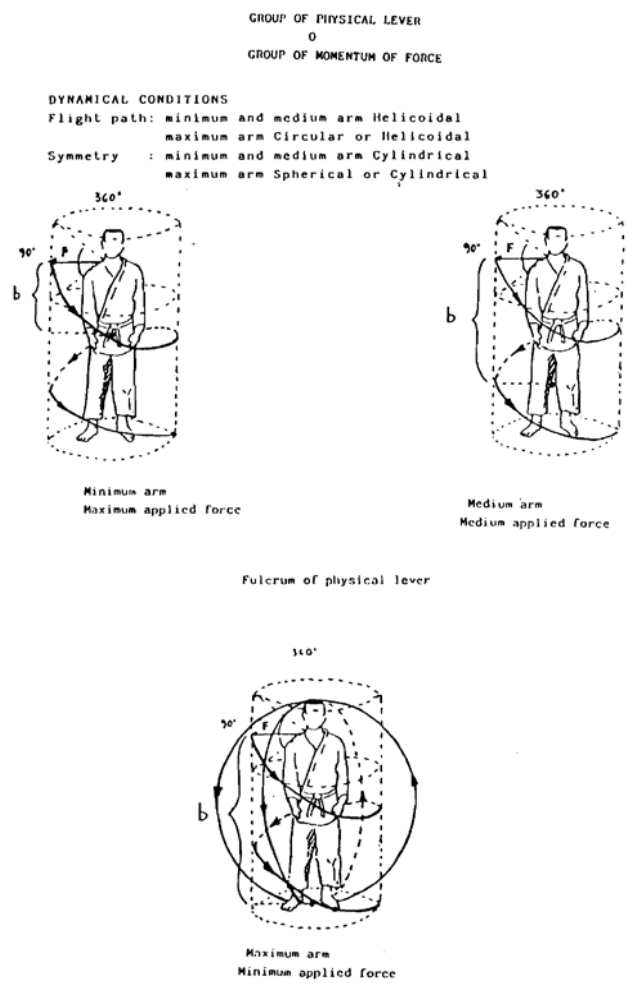

*Fig. 48: Lever Group of Throwing Techniques*

The Biomechanical Classification shows that within this group of throwing techniques considerable biomechanical similarities exist. For example, classical *ashi-guruma* and *hiza-guruma* are similar ( *opposite for Tori)* techniques applying a lever of medium arm (knee) on *uke*'s body. Interestingly, if we disregard the several possible positions of *tori*'s legs (*i.e.*, rotation of the trunk on the waist around a generic variable axis of rotation) then most throws within the Lever-type of Throwing Techniques lead to only a single basic action of *tori* ( motion of body into the transverse plane) [9,10]. see Fig 49

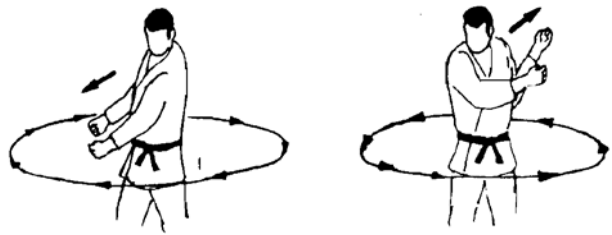

*Fig. 49: Motion within the transverse plane*

The body-sacrifice throwing techniques (*i.e.*, the in classical *Kōdōkan* jūdō terminology) so-called *sutemi-waza* must be classified as throws of the "Physical lever-type group" with maximum lever arm. In these cases the stopping point (fulcrum) is created by the friction between *uke*'s foot and the mat (*tatami*). Although such *sutemi-waza* are, energetically speaking, more favorable, the starting force (as produced by body's mass falling down) is applied with an angle greater of 45° (see the second corollary of the static analysis). To properly enable such throws and effectively throw *uke*'s body, considerable directional assistance from *tori*'s arms or legs is essential in such case the uke's body trajectory is one arc of circumference, geodetics of a spherical symmetry ( see Fig 45 c ).

Now showing the strength of the Biomechanical Analysis, we are able to group all Judo Throwing Techniques known and extensively all Judo future Throws, like : Classic, Innovative and New or Chaotic , in our biomechanics framework based on Couple and Lever Tools. These tools are the only one's able to throw the adversary's body without energy waste driving the body along specific geodetic trajectories connected to the symmetries existing into the Couple of Athletes system. In Fig 50 it is shown the Circular geodetic path of Spherical Symmetry in a Lever Throws ( Sutemi) .

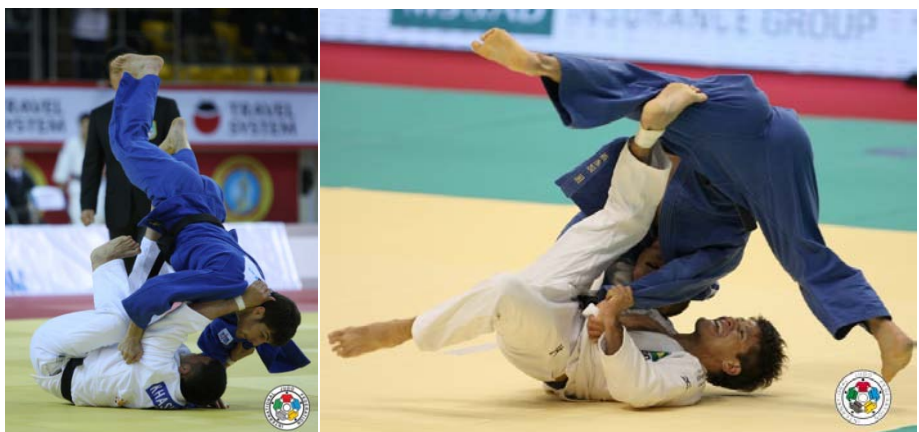

*Fig 50 Circular uke's trajectory in Sutemi*

In the following tables all the throwing techniques named in the books of Yokoyama and Oshima, Kano, Kudo, Daigo, Sacripanti, Imman, Mifune, Koizumi, Kawaishi, Inokuma and Sato are grouped under the two tools for throwing or better the basic physical principles founded in our Biomechanical revision. Tab. 1; 2
[2,5, 7, 9, 21, 34, 35, 36, 37].

| | | | |
|---|---|---|---|
| **"Couple of Forces"-type Throwing Techniques**<br><br>Couple applied by: | Arms | *Kuchiki-daoshi*<br>*Kibisu-gaeshi*<br>*Kakato-gaeshi*<br>*Te-guruma* | **All Innovative Variations of Throws and very few Chaotic Forms of Throws** |
| | Arm/s and leg | *De-ashi-barai, Ō-uchi-gari, Okuri-ashi-barai, Ko-uchi-gake, Ko-uchi-barai, Ko-soto-gake, Ō-uchi-barai, Harai-tsuri-komi-ashi, Tsubame-gaeshi, Yoko-gake, Ko-uchi-gari, Ō-soto-gake, Ko-soto-gari, Ō-uchi-gake, O-uchi-gaeshi(1)* | |
| | Trunk and legs | *Ō-soto-gari, Ō-tsubushi, Ō-soto-guruma, Ō-soto-otoshi, Uchi-mata, Ko-uchi-sutemi, Okuri-komi-uchi-mata, Harai-makikomi, Harai-goshi, Ushiro-uchi-mata, Ushiro-hiza-ura-nage, Hane-goshi, Gyaku-uchi-mata, Hane-makikomi, Daki-ko-soto-gake, Yama-arashi (Khabarelli-type throw), Uchi-Mata-gaeshi, Hane-goshi gaeshi, Harai-Goshi -gaeshi,Uchi-Mata-makikomi, Harai-makikomi, Hane-makikomi,* | |
| | Trunk and arms | *Morote-gari* | |
| | Legs | *Kani-basami* | |

*Table 1: Throwing techniques based on a Couple Tool*

| | | | |
|---|---|---|---|
| **Physical Lever-type Throwing Techniques** Lever applied by: | **Minimum Arm Lever** (fulcrum under *uke*'s waist) | *Ō-guruma, Ura-nage, Kata-guruma, Ganseki-otoshi, Tama-guruma, Uchi-makikomi, Binta-Guruma, Obi-otoshi, Soto-Makikomi, Tawara-gaeshi, Makikomi, Kata-sode-ashi-tsuri, Sukui-nage, Daki-sutemi, Ushiro-goshi, Utsuri-goshi* | **All Innovative Variation and New (Chaotic) Forms** |
| | **Medium Arm Lever** (fulcrum under *uke*'s knees) | *Hiza-guruma, Ashi-guruma, Hiza-soto-musō, Soto-kibisu-gaeshi* | |
| | **Maximum-Arm Lever** (fulcrum under *uke*'s malleolus) | *Uki-otoshi, Yoko-guruma, Yoko-otoshi, Yoko-wakare, Sumi-otoshi, Seoi-otoshi, Suwari-otoshi, Hiza-seoi, No -Waki, O-uchi-gaeshi(2) Waki-otoshi, Obi-seoi, Tani-otoshi, Suso-seoi, Tai-otoshi, Suwari-Seoi, Dai-sharin, Hiza-tai-otoshi, Hikkomi-gaeshi, Tomoe-nage, Sumi-gaeshi, Ryō-ashi-tomoe, Yoko-kata-guruma, Yoko-tomoe, Uki-waza, Sasae-tsuri-komi-ashi, Uke-nage* | |
| | **Variable Arm** (variable fulcrum from *uke*'s waist to his knees) | *Tsuri-komi-goshi, Kubi-nage, Ō-goshi, Sasae-tsuri-komi-goshi, Koshi-guruma, Ko-tsuri-komi-goshi, Ō-tsuri-komi-goshi, Sode-tsuri-komi-goshi Seoi-nage, Eri-seoi-nage, Uki-goshi, Morote-seoi-nage* | |

*Tab.2. Throwing Techniques based on a Physical Lever*

# 7. Conclusion

The biomechanical reassessment of the technical foundations of Kanō's *Kōdōkan* jūdō do not challenge or change the basic concepts of Kanō as a leading Japanese educator, but it only expands the original ideas in a clear scientific perspective. This reassessment is useful because it rectifies some minor inconsistencies that are present in the current jūdō theory.
The results obtained can be easily extrapolated in a consistent way to the contest situation.
These achievements were of great help to both jūdō teachers and coaches.

The sports aspect of jūdō was not in Kanō's mind when he developed his educational system. However, today Jūdō is known worldwide more for its sports applications than for its educational message, hence there has been a desire for developing a rigorous and rational teaching and coaching tool. The biomechanical reassessment which we have performed, allowed us to:

1. Broaden Kanō's Unbalancing concept (*kuzushi* phase) from static to dynamic situations.
2. Identify within the *tsukuri* phase the only three solutions (based on the *Jacobi Minimum Action principle [GAI]*) that enable a jūdō athlete to reduce the distances from his opponent.
3. Single out the only two basic physical tools that enable the opponent being thrown."*Couple*" and *"Lever"*
4. Understand how to make use of *Superior Kinetic Chains (SSAI)* (*kake* phase).
5. Understand how to make use of the *Inferior Kinetic Chain (ISAI)* (*kake* phase).
6. Arrange and categorize in a rational way all the possible jūdō throwing techniques that may be applied.
7. Identify the Innovative Form (ref. Roy Inman) of throws.
8. Identify the New (or Chaotic) Forms of throws.
9. Understand how Innovative and New (Chaotic) techniques tend to be created.

These results have very useful applications for rationalizing the teaching system of jūdō, both for deepening our knowledge of jūdō contests and the diverse interrelations that exist within the complex interactions that take place during such contests (*i.e.*, throws in a dynamic field).

# 8. Physical and Mathematical framework

## 8.1 General Action Invariant Theory

As previously explained, we call these movement classes aimed to shorten the distance between athletes: *General Action Invariants (GAI).*
This term covers the whole range of body movements ( rectilinear and curvilinear) intended to both reduce the distance between two opponents and to optimize one's body relative to the adversary's body position. We will demonstrate that each of these classes is defined by what is known in classical Newtonian physics as the physical *Principium of Jacobi Minimum Action*.
What is also interesting is that in biomechanics these **Actions Invariants** can be traced back to the *Hamilton –Lagrange Equation* and to the *Hamilton Action Principle*, namely:

$$GAI(q,t) = \int L(q,\dot{q},t)dt \qquad (1)$$

In the above equation, GAI= the Action (***all shortening action movement***) function of generalized position and time, and L= the Lagrange function (***essentially, the energy of the Couple of Athletes system***) depending from the generalized position coordinates, time and generalized velocity .
This Action, as first approximation, neglecting friction, considers <u>constant the external energy</u> of the system (***gravitational field***) after integration, it is possible to write:

$$GAI(q,t) = W(q) - Et = L(q,\dot{q}) \qquad (2)$$

*if we perform a little variation of the shortening action movement*

*the total energy of the system is constant and its variation is obviously zero*

$$\delta GAI(q,t) = \delta \int_{t_1}^{t_2} L(q,\dot{q})dt = \delta \int_{t_1}^{t_2} \left[ \frac{\partial L}{\partial q} - \frac{d}{dt}\left(\frac{\partial L}{\partial \dot{q}}\right) \right] \delta q \, dt = 0 \Rightarrow \qquad (3)$$

*then the mechanical energy is constant*

$$\frac{\partial L}{\partial q} - \frac{d}{dt}\left(\frac{\partial L}{\partial \dot{q}}\right) = 0 \qquad (4)$$

*if the system is not conservative, we must write*:

$$\frac{\partial E}{\partial q} - \frac{d}{dt}\left(\frac{\partial E}{\partial \dot{q}}\right) + Q = 0 \qquad (5)$$

According to <u>Hamilton's Principle</u>, the true evolution of GAI(q,t) is an evolution for which the shortening action movement is <u>stationary</u> (a minimum, a maximum, or a saddle point). Normally, during a jūdō action one pursues the minimum solution, *i.e.* the so-called the Principle of Minimum Action.
Generally, this applies to a conservative field; if, however, we consider a non-conservative field, then it also becomes necessary to incorporate a factor Q (equation 5), *i.e.* the heat emitted, which is part of the global balance; in such a situation it becomes impossible to establish such a minimum for that shortening action movement.

In jūdō competition though, such shortening action movements occur so fast one after another (0.4-0.6 sec) [30, 31, 32, 33 ] that the interaction (complete throws) ( medium time 0.8 -1.0 sec) [30, 31, 32, 33] within the couple of athletes system could be judged, with very good approximation, adiabatic (without thermal variation during the action); for this reason it is then appropriate to calculate these fast shifts (shortening action movements GAI) that occur in competitive jūdō as applications of the principle of minimum action, hence the name of **General Action Invariants**

Jigorō Kanō's jūdō principle of maximum efficiency with minimum effort considering the trajectories, in jūdō biomechanics is explained by the *Jacobi form of the least action* principle, which in general formulation is:

$$\Delta \int_{\rho_1}^{\rho_2} \sqrt{E + W(\rho)} \, d\rho = 0 \qquad (6)$$

*with $E =$ kinetic energy*

*and $W =$ potential energy*

These results are valid with a good approximation in two dimensions, such as projections of the Center of Mass (COM ) of an athlete on the *tatami*. As to the more real and complex 3-dimensional problem, it is possible in terms of a mathematical expression to also find a general solution for the "shortening of distance action", which we call *General Action Invariants (GAI).* [19].

Considering the infinite situations that arise during a jūdō competitive match, providing an analytical solution presents an extremely difficult challenge. However, the geometrical approach does shed some light on this very complex problem. This solution may be acceptable if one considers the *Couple of Athletes*-system for a stationary situation which also is relevant for application of any physical lever-based jūdō throw during which athletes have to momentarily come to a standstill.

Very often and with very good approximation, the type of motion performed by a *Couple of Athletes* which applies throwing techniques that range within the couple of forces group may be considered as being nearly uniform, both linear and circular with a mean velocity ranging around 0.3 m/s [15, 16]. If, in these cases we properly change the reference system ("*Galilean Relativity*"), we are able to reduce this system to a stationary position.
This is the reason for which the static analysis will fit almost well the dynamic situation.

For the other two classes of *General Action Invariants* based on rotations, some interesting derivations can be made from the Poinsot geometrical description of a free-forces motion of a body. in such case in fact the motion is like a rolling of the body inertial ellipsoid (without slipping) on a specific plane, remembering that the curve traced out by the point of contact on the inertial ellipsoid is called polhode, while the curve on the plane is called herpolhode. When considering a jūdō athlete, the body could be regarded as having cylindrical symmetry and the inertial ellipsoid becomes a revolution ellipsoid, in such case the polhode is a circle around the athlete's axis of symmetry, and the herpolhode on the plane (*i.e.* the tatami) represents a circle.

The limitation is that these results into the "athletes reference system" [10] are real only in the case of free force motion; it is therefore applicable if we consider the motion of the couple as one entity or as a whole, which is the motion called of free body.
But in reality in jūdō there exist INTERACTIONS that are going on and that are being produced between the couple into the couple, such as their pushing/pulling forces as well as the friction forces that are acting on them. This problem is almost not analytically solvable for this degree of complexity.

## *8.2 Interaction*

The general physics-biomechanical model for jūdō throws (*i.e.*, what in technical language in physics is known under the term '***Interaction***') is similar to the "two- bodies problem with collision" in classical physics. We treat before the two body problem. The general potential that describes this interaction has the general exponential form:

$$V' = r^{-\alpha} + r^{-2\alpha} \tag{7}$$

It is well known and possible to demonstrate [14] that the common part of interaction can be described by the curves family showed in a generalized *Morse's Potential*:

$$V = D\left(e^{-2\alpha(r-r_0)} - e^{-\alpha(r-r_0)}\right) \tag{8}$$

V' in (7) is a particular expansion of the expression (8).
The specification of a general form of interaction potential (8) is able to provide us a lot of useful information

1. $r_o$ is the equilibrium distance (*i.e.*, the gripping distance in jūdō).
2. D is the mechanical potential energy in the equilibrium point ($r_o$), which is equal to the mechanical mean energy as expressed in terms of oxygen consumption as $\eta O_2$ (**"*Sacripanti Relationship*"**) [28].
3. It is possible to evaluate the constant α expanding the potential (8) near the minimum.

The connection with the harmonic term of expansion then becomes:

$$D\alpha^2 (r-r_0)^2 = E_c \quad or \quad \alpha = \frac{1}{L}\sqrt{\frac{E_c}{D}} \tag{9}$$

To know the potential let us go back to the Algebraic expression of force:

$$F = ma = \frac{dV}{d\alpha} = 2\alpha D\left(e^{-\alpha r} - e^{-2\alpha r}\right) \tag{10}$$

Singling out the common part of the interaction as a "two- bodies problem in a central field" allows us to make use of an important given from classical physics about the mean time value of several variables (*i.e.*, *Virial's Theorem*). *Virial's* (or, *'Clausius'*) *Theorem* applies to both motion and interaction, and assures: if the generalized force F is a sum of friction and central forces, that the mean kinetic energy of the system in time is independent from friction forces. Most interesting is the analysis of motion of athletes tied together by grips ( bounded motion in physical term): one athlete around his opponent. In particular it is possible to show a powerful theorem that gives us information about the central forces that lead to *closed orbits.*

For every given this will occur if the equivalent potential V' will have an "extremal" point (maximum or minimum) at a determined point $r_0$, and if the Energy E is just equal to V'($r_0$). In $r_0$, if V' has an "extremal" point (this means that F' is equal to zero), then the force in $r_0$ will be as considered by the following equation:

$$F' = F + mr\dot{\theta}^2 = F + \frac{l^2}{mr^3} \qquad (11)$$

The information as expressed here, indicates that the orbit will be closed (*i.e.*, "pseudo circular") for attracting forces. The energy can also be evaluated as:

$$F(r_0) = -\frac{l^2}{mr_0^3} \qquad (12)$$

$$E = V(r_0) + \left(\frac{l^2}{2mr_0^2}\right) \qquad (13)$$

When evaluating the situation for either maximum or minimum values of V', according to the second derivative, if the curve is concave up (*i.e.*, U-shaped) and V' positive, there will a bounded orbit that is stable (closed circular); when, on the other hand, V' is concave down (*i.e.*, reverse U-shaped) the orbit is unstable and unbounded. The *Bertrand Theorem* assures us that, normally, closed orbits are possible for forces including a dependent variable $1/r^2$. However, if the energy is greater than the energy of a circular orbit, the orbit under certain conditions could be with either open or closed recurrence.
For example, developing the force, in expansion of Taylor F(r), will be a function of a parameter β that arises from the question that the motion is harmonic in $1/r_0$.

$$\frac{1}{r} = \frac{1}{r_0} + A\cos\beta\theta \qquad (14)$$

The amplitude A is connected to the Energy deviation from the circular orbit; β comes from the Taylor series expansion of force about the circular orbit, around $r_0$.
Substituted into a force equation, this gives:

$$\beta^2 = 3 + \frac{r}{f}\frac{df}{dr}\bigg|_{r=r_0} \qquad (15)$$

If the radius r goes around the plane, the quantity 1/r will go through β cycles of its oscillation. This leads to the following important consideration. It is well known in physics that if the attracting force is represented by the Law of Power expressed as $F=-kr^{\alpha}$, then only $\alpha = -2$ and $\alpha = 1$ will represent closed orbits.

The projection of the two Center of Mass (COM) trajectories on the *tatami* will be a closed planar orbit. Fig.51 (a).

But the hypothesized "gravitational model" between the two athletes not satisfy a good representation of the trajectories on the *tatami* during rotational shifting movements (closed orbits) because trajectories can't intersect themselves the presence of the gripping arms hampers such situation. Luckily *Sundman's Theorem* provides the link between the solutions for nonlinear (Newton's Law) and linear (Hooke's Law) problems. Each orbit under Newton's Law is an image of a Hooke's Law orbit under the transformation $z \leftrightarrow z^2$.
Hooke's law is more relevant for the push/pull action really developed by grips in Judo motion/competition.

Remembering that the Couple of Athletes system is, neglecting friction, a free body (no external force on it) it could be modeled as two masses connected, not by the gravity force but rather by springs (gripping arms) with no gravity force (because the mat stops gravity action [free body]) to which should be added the friction between the feet and mat (under push/pull forces actions). The projections of the two Center of Mass (COM) trajectories on the *tatami* will be two closed planar orbits that never intersect each other. These orbits produced by this duality are shown in the next figure for all the two cases considered. The first trajectory is the result of a model that considers a gravitational force between the athletes, the second one's is the Hooke trajectories in a gravitational field at 1.G ; the third are the trajectories of Couple of Athletes model that considers a spring-like force at 0.1 G (more real) between athletes.
The red path and the blue one's represent the gripped Athletes COM Projections trajectories during a rotational phase of the contest. Fig.51

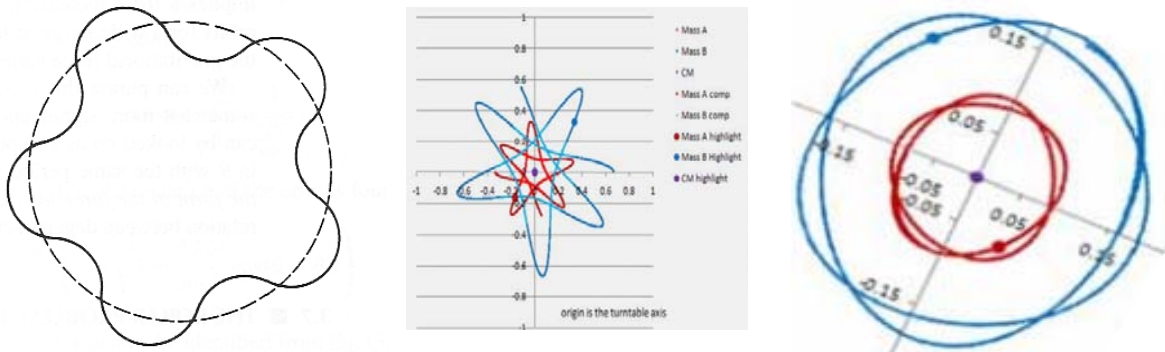

*Fig.51: (a) Newton's Orbit ($F=1/r^2$) of the projection of the COM of the attacker around the opponent with Energy deviating slightly (greater) from a circular orbit β=5; (b) & (c): Same orbits according to Hooke's Law ($F= -Kr$)[grips] at 1G (wrong) and 0.1G very similar to real competition condition. ( from NASA Experiments)*

However the system (Couple of Athletes) is isolated from external forces, and on it act only friction and the internal random push-pull forces, then the system follows a *Langevin-type Equation*, as we have already demonstrated and evaluated in the *first Sacripanti model* [14]:

$$F = -\mu v + mv \sum_j (\pm 1)_j \delta(t - t_j) = F_f + F_r \qquad (16)$$

In fact there are more hidden complexities, such as, for example, the geometrical adaptation of the structure of muscles involved which alter the amount of contractile force; this, in fact makes the problem analytically impossible to resolve. In any way in the *Second Sacripanti Model* it was already shown that the Couple of Athletes system in jūdō competition moves according to a fractional Brownian motion. Nevertheless, we have already discussed this in detail elsewhere [7, 11, 14-18]

## 8.3 Throws Collision theory

Interaction was already solved in [16 and 28], however we will treat a lesser known aspect of Bodies interaction by Judo Throwing techniques, but not less important. " Collision among Athletes bodies parts".

As already demonstrated before, the collision between athletes is a very important part of the new concept of breaking symmetry and throw connected to the unbalance action in dynamic condition.

In short athlete breaks the competitor symmetry this action shifting the COM inside the body slowing down the adversary that is unable to shift away fast, then simultaneously starts the tsukuri phase that end with the body collision parts, starting the *kake* phase.
The collision will be different in function of tool utilized to throw, namely Couple or Lever.
In the following paragraphs we will analyze specifically this two kind of collisions.

### 8.3.1 "Couple " Throws
In this case of collision produced by the Couple tool, mainly the collision happens between two legs of athletes, or some specific parts of them.
In such throwing Judo actions, the collisions are the main movement of inferior dynamic chains in the Couple groups (arms and leg, trunk and leg). The collisions as applied in these techniques are "off-center type of collision" with friction. The **off-center collision** does not normally follow the connecting line of the athlete's center of gravity; hence there is a torque present. The *uke*'s body begins to rotate around his COM, in this movement being also assisted by the other side of the couple application.

Because the athletes' bodies represent macroscopic bodies we can use the goal of *penalty models* to study their collision action. In such case we start studying the deformation of colliding bodies instead of studying the evolution of velocity before and after the collision.
However, since this deformation is almost impossible to mathematically model, one has to rely on the common hypothesis that the force is described by an equation that takes into account the occurrence of springs compression (muscles deformation ) plus friction:

$$F = -\lambda x_i - \mu v_i \qquad (17)$$

This force represents the two bodies' deformation caused by their collision with the involved friction.
The two parameters are connected by the relationship of the *restitution coefficient*:

$$e = 1 - \alpha v \qquad (18)$$

*that let us evaluate the friction coefficient μ in function of elastic constant λ and restitution coefficient e*

$$\mu = \frac{3}{2}\alpha\lambda = \frac{3}{2}\left(\frac{1-e}{v}\right)\lambda \qquad (19)$$

Without going into the mathematical difficulties that lie beyond this example, the equation (17) and the collision as described are connected to the following differential equations:

$$F^l = ma \qquad (20)$$

*linear force*

$$N = I\ddot{\omega} + \dot{\omega} \wedge (I, \dot{\omega}) \qquad (21)$$

*rotational component*

$$N = g(F) \qquad (22)$$

*linear and rotational forces connection*

$$F_t = f(x,v) + F^l + N \qquad (23)$$

*all forces connection*

Solving these equations let us to know interesting aspects of the two limbs collision phenomena, during the application of a Judo Throwing Technique with a "Couple" tool.

### 8.3.2 "Lever" Throws

This collision could be analyzed considering it as an inelastic non-central collision with the uke's body at rest.

Normally, for the Lever group of throwing techniques, the actual collision can occur at a stopping point. However, in real contest situations before the stopping point is applied, the Dynamic Unbalance Action ends with a collision between *tori*'s and *uke*'s bodies and not limbs. That happens, very often, in the Physical lever throws: group of "minimum arm", because in such throws opponent's bodies join.

The *tori*'s body then with a momentum **p** = m**v** collides with the *uke*'s body technically at rest **p** = 0. After the collision, *tori*'s body moves with a decreasing momentum **p'** = m**u**, while *uke*'s undergoes a recoil momentum **p"** = m **u'**.
The collision process cannot be entirely expressed by the energy and momentum laws, since there are only 4 equations for calculating the 6 components of the final momentum.
The end points of **p'** lies on the **momentum sphere** with a radius r= $p' \, m_u / (m_T + m_U)$, where the center of this sphere divides the momentum **p'** according to the ratio of the two bodies' masses. In our case the tool Lever applies an inelastic collision with friction, then the radius of the momentum sphere changes, while the center remains in place.
The radius, however, increases because during the throw's inelastic collision the total energy decreases $\Delta E < 0$.

In real situation this inelastic collision with friction between the two bodies occurs at lesser velocities than two limbs, this let us easy understand why the two athletes' bodies oftentimes remain practically connected into the trajectory after throw collision..

# 9. Acknowledgements

The author acknowledges for permission to use their photographs: David Finch and Tamás Zahonyi (IJF Archive) Courtesy of the IJF President, Marius Vizer, Professor Carl De Crée for the valued help and assistance with the English and Japanese grammar, Rodney Imamura and Stanislav Sterkowicz, for their reviews during the preparation of this manuscript.